\documentclass[journal ,12]{IEEEtran}
\usepackage{colortbl}
\usepackage[table]{xcolor}

\usepackage{tikz}
\usepackage{algorithm}
\usepackage{algorithmicx}
\usepackage{algpseudocode}
\usepackage{url}
\usepackage{float} 
\usepackage{amsmath,amsfonts,amssymb,mathrsfs,amsthm,amsbsy,graphicx,color,bm,algpseudocode,balance,mathtools,xfrac,nccmath,siunitx,cite}
\usepackage{multirow}

\newtheorem{rem}{Remark}

\usepackage[font=scriptsize,labelfont=bf]{caption}

\usepackage{pgfplots}
\pgfplotsset{compat=newest}
\usetikzlibrary{plotmarks}
\usetikzlibrary{arrows.meta}
\usepgfplotslibrary{patchplots}
\usepackage{grffile}
\usetikzlibrary{plotmarks}

 \usepackage[utf8]{inputenc}
\usepackage{booktabs, caption, makecell,xspace}

\usepackage{threeparttable}

\newcommand{\bbc}{\text{BiBC}\xspace}
\newcommand{\abc}{\text{AmBC}\xspace}

 \theoremstyle{definition}
 \makeatletter
\newcommand{\printfnsymbol}[1]{%
  \textsuperscript{\@fnsymbol{#1}}%
}

\usepackage{pifont}
\newcommand{\cmark}{\ding{51}}%
\newcommand{\xmark}{\ding{55}}%
\newcommand{\lmark}{\textbf{--}}

\makeatother
\begin{document}

\title{Cell-Free Bistatic Backscatter Communication: Channel Estimation, Optimization, and Performance Analysis}

\author{Diluka  Galappaththige\textsuperscript{\textasteriskcentered}\thanks{\printfnsymbol{1}D.  Galappaththige and F. Rezaei contributed equally to this work.}, \IEEEmembership{Member, IEEE}, Fatemeh Rezaei\printfnsymbol{1}, \IEEEmembership{Member, IEEE},   Chintha Tellambura, \IEEEmembership{Fellow, IEEE,}   Amine Maaref, \IEEEmembership{Senior Member, IEEE}
\thanks{D. Galappaththige, F. Rezaei, and C.~Tellambura with the Department of Electrical and Computer Engineering, University of Alberta, Edmonton, AB, T6G 1H9, Canada (e-mail: \{diluka.lg, rezaeidi, ct4\}@ualberta.ca).  \\
\indent A. Maaref is with Huawei Canada, 303 Terry Fox Drive, Suite 400, Ottawa, Ontario K2K 3J1 (e-mail: amine.maaref@huawei.com).}
\vspace{-5mm}}


\maketitle

\begin{abstract}
This study introduces and investigates the integration of a cell-free architecture with bistatic backscatter communication (\bbc), referred to as cell-free \bbc or distributed access point (AP)-assisted \bbc, which can enable potential applications in future (EH)-based Internet-of-Things (IoT) networks. To that purpose, we first present a pilot-based channel estimation scheme for estimating the direct, cascaded, forward channels of the proposed system setup. We next utilize the channel estimates for designing the optimal beamforming weights at the APs, reflection coefficients at the tags, and reception filters at the reader to maximize the tag sum rate while meeting the tags' minimum energy requirements. Because the proposed maximization problem is non-convex, we propose a solution based on alternative optimization, fractional programming, and Rayleigh quotient techniques. We also quantify the  computational complexity of the developed algorithms. Finally, we present extensive numerical results to validate the proposed channel estimation scheme and optimization framework, as well as the performance of the integration of these two technologies. Compared to the random beamforming/combining benchmark, our algorithm yields impressive gains.  For example, it achieves  $\sim\qty{64.8}{\percent}$ and $\sim\qty{253.5}{\percent}$ gains in harvested power and tag sum rate, respectively, for \qty{10}{\dB m} with \num{36} APs and \num{3} tags.
\end{abstract}



\IEEEpeerreviewmaketitle
\section{Introduction}
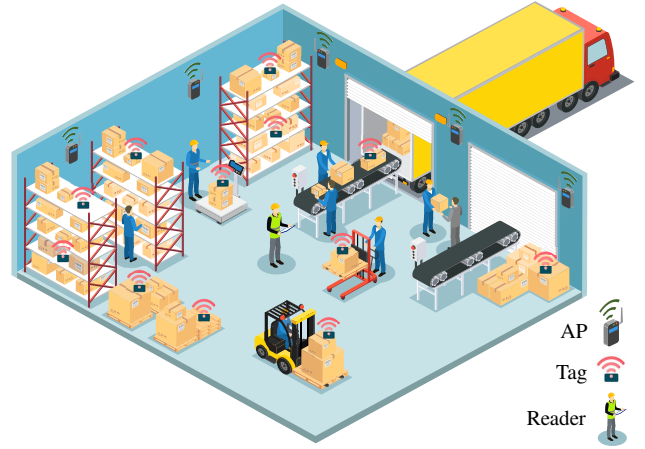
\begin{figure}[!t]\centering \vspace{0mm}
    \def\svgwidth{240pt} 
    \fontsize{8}{8}\selectfont 
    \graphicspath{{Figures/}}
\begingroup%
  \makeatletter%
  \providecommand\color[2][]{%
    \errmessage{(Inkscape) Color is used for the text in Inkscape, but the package 'color.sty' is not loaded}%
    \renewcommand\color[2][]{}%
  }%
  \providecommand\transparent[1]{%
    \errmessage{(Inkscape) Transparency is used (non-zero) for the text in Inkscape, but the package 'transparent.sty' is not loaded}%
    \renewcommand\transparent[1]{}%
  }%
  \providecommand\rotatebox[2]{#2}%
  \newcommand*\fsize{\dimexpr\f@size pt\relax}%
  \newcommand*\lineheight[1]{\fontsize{\fsize}{#1\fsize}\selectfont}%
  \ifx\svgwidth\undefined%
    \setlength{\unitlength}{400.73538208bp}%
    \ifx\svgscale\undefined%
      \relax%
    \else%
      \setlength{\unitlength}{\unitlength * \real{\svgscale}}%
    \fi%
  \else%
    \setlength{\unitlength}{\svgwidth}%
  \fi%
  \global\let\svgwidth\undefined%
  \global\let\svgscale\undefined%
  \makeatother%
  \begin{picture}(1,0.70813277)%
    \lineheight{1}%
    \setlength\tabcolsep{0pt}%
    \put(0,0){\includegraphics[width=\unitlength]{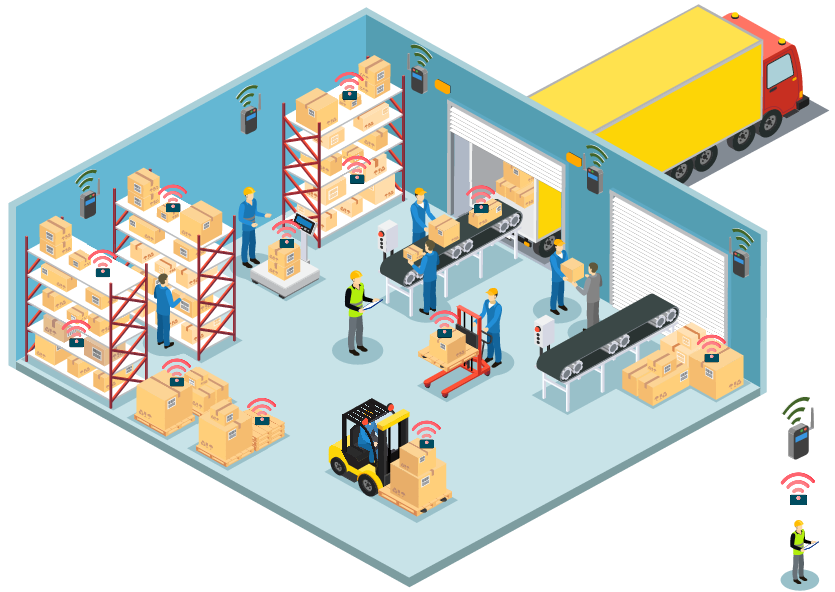}}%
    \put(0.82255202,0.03489126){\color[rgb]{0,0,0}\makebox(0,0)[lt]{\lineheight{1.25}\smash{\begin{tabular}[t]{l}Reader\end{tabular}}}}%
    \put(0.8691627,0.10701334){\color[rgb]{0,0,0}\makebox(0,0)[lt]{\lineheight{1.25}\smash{\begin{tabular}[t]{l}Tag\end{tabular}}}}%
    \put(0.87611158,0.17250681){\color[rgb]{0,0,0}\makebox(0,0)[lt]{\lineheight{1.25}\smash{\begin{tabular}[t]{l}AP\end{tabular}}}}%
  \end{picture}%
\endgroup%
 
    \caption{A warehouse use case of cell-free \bbc network.}\label{fig_systemGeneral}
    \vspace{-0mm} 
\end{figure}

With the rapid deployment of fifth-generation (5G) and beyond 5G (B5G) communication networks, energy harvesting (EH)-based Internet-of-Things (IoT) is becoming an extremely active research area, and the third-generation partnership project (3GPP) has launched a new study item \cite{Passive_IoT_design, Huawei_ambient, Huawei}. These research items identify the following essential aspects and aims of EH-based IoT networks: 
\begin{itemize}
    \item Use cases such as identification, tracking, monitoring, actuating, and sensing for applications in logistics, transportation, healthcare, and others.
    \item Exploring public/private networks, indoor/outdoor environments, macro/micro/pico cells, cell-free connectivity to user equipment (UEs)  with or without relay/UE assistance, and frequency bands (both licensed and unlicensed).
    \item Establishing EH techniques, connectivity requirements, and positioning accuracy.
\end{itemize}
These challenges open a vast array of research questions. Very few of those have been explored  \cite{Galappaththige2022, Rezaei2023}. Inspired by them, we consider the problem of supporting the energy needs of multiple tags over a large coverage area such as a warehouse with a bistatic backscatter (\bbc) network of dedicated access points (APs)/radio frequency (RF) sources (see Fig. \ref{fig_systemGeneral}). Furthermore, these inexpensive backscatter tag-assisted IoT networks have numerous applications, including logistics, inventory management, warehousing, manufacturing, energy industry, healthcare, agriculture, aerospace and defense, farming, retail, sports, and many more. The potential market will grow at an exponential rate; for example, parcel volume in China will reach \num{163.4} billion by \num{2025} and \num{220}-\num{262} billion globally by \num{2026} \cite{Huawei, Passive_IoT_design, Huawei_ambient}. These possible use cases offer significant market opportunities.

In scenarios like those mentioned, backscatter tags are favored due to their cost-effectiveness and minimal energy consumption (a few nW-$\mu$W). They reflect RF signals from the RF source to transmit data to a reader, whether it is a dedicated or cooperative one \cite{Galappaththige2022, Rezaei2023, HoangBook2020}, thus optimizing spectrum utilization and reducing the need for additional spectrum allocations. Tags find utility in IoT networks, enhancing both spectrum and energy efficiency \cite{Galappaththige2022, Rezaei2023, HoangBook2020}. Among the three backscatter network types (monostatic, bistatic, and ambient) \cite{Galappaththige2022, Rezaei2023}, \bbc systems excel in applications like warehouses. These systems employ dedicated RF sources, single or multiple, to power tags and enable backscatter modulation \cite{Galappaththige2022, Rezaei2023}. In the context of \bbc, dedicated RF sources offer advantages over ambient signals, including predictability, reduced interference, system control, and knowledge of source signal parameters. These advantages can be harnessed to optimize system coverage and performance \cite{Galappaththige2022, Rezaei2023}.

\subsection{Technical Challenges in \bbc} \label{sec_challenges}

Despite their wide application potential, several critical challenges constrain the performance of \bbc systems.
\begin{enumerate}
    \item 
First, the dedicated RF sources (power beacons) in \bbc are primarily designed to provide energy wirelessly to tags.   Power beacons offer localized coverage, providing energy within a specific area.  Tags have   EH  circuits to convert the dedicated RF signals to direct current \cite{Galappaththige2022}. Specifically, the performance of the EH circuits depends on the activation threshold, typically around \qty{-20}{\dB m} \cite{GAOtags} or more of incident RF power. Hence, depending on the proximity to the power beacon, tags may not receive adequate power to activate EH and perform data backscattering. This causes energy outages at the tags and short activation ranges (a few meters).
Hence, a single power beacon supporting multiple tags limits the tag performance while increasing the probability of an energy outage. Consequently, supporting a large number of tags across a broader area may need multiple power beacons. 

\item Second, as tags backscatter their data, tag signals experience double path losses and deeper fading, leading to short communication distances ($\leq\qty{6}{\m}$), low data rates ($\leq \qty{1}{bps/\Hz}$), and low reliability \cite{Galappaththige2022,Rezaei2023}. While connecting multiple tags is vital in applications,  mutual interference caused by concurrent tag transmissions degrades the network performance. Hence, sophisticated signal processing techniques, e.g., beamforming design, interference cancellation schemes, and multiple access schemes are essential to increase reliability while establishing massive connectivity. 

\item Third,  optimal beamforming design, effective suppression of the direct link interference from the power beacon, and accurate data decoding of tags critically depend on the availability of perfect channel state information (CSI). However, the inherent limitations in the power and processing capabilities of tags, which simply backscatter incident RF signals, introduce substantial challenges to the channel estimation process. This complexity is further compounded in scenarios involving multiple power beacons and multiple  tags.  Thus, sophisticated methods are required for precise estimation of the direct channel ($\mathbf{h}_{0,m}$),  forward channel ($f_{k,m}$), and the cascaded channels ($f_{mk} \mathbf{g}_k$) -- Fig. \ref{fig_system_fig}.  
\end{enumerate}

\subsection{Motivation  and Our Contributions} Inspired by the aforementioned challenges in \bbc, as well as significant gaps in the literature to address the needs of EH-based IoT networks, we propose supporting multiple tags over a large coverage area such as a warehouse with a \bbc network of dedicated distributed APs, i.e., a cell-free  \bbc system.

The basic concept behind cell-free networks is that a large number of spatially distributed APs serve multiple users on the same time-frequency resources. It thus minimizes transmission distances while increasing coverage, resulting in increased macro-diversity and favorable propagation \cite{Ngo2018, Nayebi2017, Diluka2019}. Hence, by shortening AP-tag distances, the cell-free architecture can alleviate the limited energy availability and adverse effects of path loss in \bbc \cite{Zhuoyin2023, Jia2021,  Zhuoyin2021, Han2017}. Furthermore, a central processing unit (CPU) coordinates the APs that are connected to it via a front-haul/back-haul link, allowing the APs to serve users in the area collaboratively. The distributed APs in the \bbc network can thus employ the jointly designed beamforming weights to deliver as much power to the tags as possible while minimizing inter-tag interference to support a large number of tags.

Integrating cell-free architecture with \bbc poses several critical technical challenges, such as (i) channel estimation (ii) enabling multi-tag transmission, (iii)  AP beamforming strategies, (iv) tags' reflection coefficients/power allocation, (v) designing reader reception filters, (vi) minimum energy for tag activation, (vii) coverage, (vii) coexistence with conventional cellular networks, and more. Although none of these challenges have been thoroughly investigated, \cite{Zhuoyin2021, Zhuoyin2023, Han2017, Jia2021} address some of them (Section \ref{sec_prev_cont}). Hence, to address these challenges while filling a significant gap in the literature, we propose a generalized cell-free \bbc system (Fig.~\ref{fig_system_fig}). 

In particular, in the channel training phase, APs operate in full-duplex (FD) mode and use a TDMA scheme, i.e., an on-off switching protocol that is controlled by the CPU, to estimate the direct (APs-reader), cascaded (APs-tags-reader), and forward (APs-tags) channels using specially designed pilot sequences. During the data transmission, the APs operate in half-duplex (HD) mode and cooperate to service the tags in the area by beamforming to improve the tags' rate performance at the reader while ensuring the tags' minimum energy requirements for tag activation. We thus design the optimal AP beamforming weights, tags' reflection coefficients, and reader's combing filters to maximize the tags' sum rate, while meeting the EH requirements at the tags.  We also take into account the effect of estimated CSI while designing the aforementioned AP beamforming weights, tag reflection coefficients, and reader combing filters. 

The main contributions of this paper can be summarized as follows:
\begin{enumerate}
    \item Using a specifically developed pilot sequence presented in \cite{rezaei2023timespread}, we propose a method to estimate the channels of the cell-free \bbc system, including direct, cascaded, and forward channels. During this phase, an on-off switching protocol is adopted and the APs operate in FD mode. We thus derive least squares (LS) and minimum mean square error (MMSE) estimators for these channels.

    \item Following that, we formulate the tag sum rate maximization problem by optimizing the beamforming weights at the APs, reflection coefficients at the tags, and receiver combiner at the reader, while also maintaining the EH requirements at the tags. The formulated optimization problem also accounts for the estimated CSI and CSI estimation errors. 

    \item The resulting maximizing problem has a non-convex objective function and constraints. Convex problems, on the other hand, can be solved efficiently with polynomial convergence time using readily available and widely accessible convex solvers. Therefore, we use the alternating optimization (AO) technique to decouple the non-convex problem into three sub-problems. Using this method, we develop solution algorithms for these sub-problems utilizing fractional programming (FP) and Rayleigh ratio quotient approaches.

    \item We show that the proposed channel estimation scheme and optimization frameworks improve backscatter tag performance significantly in cell-free \bbc. We also investigate the algorithms' robustness to CSI errors and the computational complexities.  

    \item We also investigate the effects of fixed reflection coefficients at the tags. In particular, while these tags have lower performance than reconfigurable tags, i.e., tags with variable reflection coefficients, they could be a low-cost networking solution for particular applications.

    \item Finally, we provide comprehensive numerical examples that demonstrate the performance of the proposed channel estimation scheme and cell-free \bbc network using the proposed solution.
\end{enumerate}

Before proceeding to the technical contributions, we provide an overview of related works that attempt to address the \bbc problems outlined in Section \ref{sec_challenges}.

\subsection{Previous Contribution on \bbc Systems} \label{sec_prev_cont}
While no prior literature thoroughly addresses all the \bbc challenges, previous studies \cite{Jia2021, Han2017, Hua2021, Kaplan2022, Tao2021}  tackled some of them. In particular, \cite{Jia2021, Han2017} utilize a set of distributed RF sources to extend the coverage distance, thereby overcoming the tag's power limitations. Reference \cite{Jia2021} specifically investigates the placement of power beacons to maximize the coverage distance taking into account the outage probability, while \cite{Han2017}  explores tag-to-tag communication and investigates the performance of network coverage and capacity. These works, however, consider single antenna nodes and overlooked the importance of beamforming designs.

On the other hand,  \cite{Hua2021, Kaplan2022,Tao2021} focus on the performance studies of a \bbc with a single tag and a single RF source. In particular, \cite{Hua2021} modifies the tag architecture to adjust the circuit load impedance of the tag to extend the communication range and lower the integrated circuit (IC) power consumption. The works \cite{Kaplan2022, Tao2021} develop interference cancellation schemes to suppress the direct link interference from the RF source, thereby improving the tag's rate. \cite{Kaplan2022} uses beamforming to nullify the direct link interference, while \cite{Tao2021} explores the joint design of waveform at the RF source and coding at the tag to mitigate the direct link interference. To further support multiple tags, \cite{Sacarelo2021} and \cite{Yang2023} respectively adopt non-orthogonal multiple access (NOMA) and time division multiple access (TDMA) schemes. \cite{Sacarelo2021} maximizes the minimum tag rate by optimizing transmit beamforming at the source, receive beamforming at the reader, and reflection coefficients at the tags. Whereas, \cite{Yang2023} minimizes the transmit power level/energy consumption at the emitter by jointly optimizing the transmission time slot duration and tag reflection coefficients. Reference \cite{Yang2020} also investigates the emitter power allocation and energy consumption considering a separate reader for each tag.

Leveraging the concept of distributed RF sources,  the authors in  \cite{Zhuoyin2021} and \cite{Zhuoyin2023} further explore beamforming design at distributed cellular APs serving a cellular user and a tag. They also consider channel estimation through pilot transmissions from the cellular APs. Channel estimation has been also investigated in \bbc using  a pilot-based two-phase channel estimation protocol \cite{Liao2022, Kaplan2022}.  However, no method exists for estimating the channels of a generalized \bbc network comprising multiple RF sources, multiple tags, and a reader. 

The most relevant works are summarized in Table \ref{comparison}, highlighting the unique contribution of this paper.

\begin{table*}[]
\caption{Summary of related works.}
\label{comparison}
 \begin{center}
\begin{threeparttable}
\begin{tabular}{|l|l|lll|l|l|l|l|l|}
\hline
\multirow{2}{*}{Conf.} & \multirow{2}{*}{Ref.} & \multicolumn{3}{c|}{Setup}                          & \multirow{2}{*}{Objective} & \multirow{2}{*}{\begin{tabular}[c]{@{}l@{}} EH  \\Constraint \end{tabular} } & \multirow{2}{*}{Variables$^\dagger$} & \multirow{2}{*}{Methodology} & \multirow{2}{*}{\begin{tabular}[c]{@{}l@{}} CSI  \\Estimation \end{tabular} } \\ \cline{3-5}
                               &                            & \multicolumn{1}{l|}{APs} & \multicolumn{1}{l|}{Tags} & Reader &                            &                                &                            &                              &                                 \\ \hline \hline
     \abc                          &    \cite{Zhuoyin2023}                        & \multicolumn{1}{l|}{$M\ge1$}  & \multicolumn{1}{l|}{$K=1$}  &  $L=1$&              Tag's rate              &      \xmark                           &            $\mathbf{w}$               &     SCA                         &      \cmark                           \\ \hline
\multirow{5}{*}{\bbc}              &    \cite{Jia2021}                        & \multicolumn{1}{l|}{$M\ge1$}  & \multicolumn{1}{l|}{$K\ge1$}  & $L=1$  &  Coverage distance                          &     \xmark                           &   AP location                          &  \lmark                            &    \xmark                             \\ \cline{2-10} 
 &   \cite{Sacarelo2021}$^\ddagger$                         & \multicolumn{1}{l|}{$M=1$}  & \multicolumn{1}{l|}{$K\ge1$}  & $L\ge 1$  &   Minimum Tag rate &     \xmark  &  $\mathbf{w}, \mathbf{u}, \boldsymbol{\alpha}$                          &  AO \& SCA                              &  \xmark                               \\ \cline{2-10}
                               &   \cite{Yang2023}                         & \multicolumn{1}{l|}{$M=1$}  & \multicolumn{1}{l|}{$K\ge1$}  & $L=1$  &    Transmit power                        &  \cmark                               &  $p_t, \alpha$                          &  SCA \& BCD                             &  \xmark                               \\ \cline{2-10} 
                               &    \cite{Yang2020}                        & \multicolumn{1}{l|}{$M=1$}  & \multicolumn{1}{l|}{$K\ge1$}  &$L=1$  &    Energy efficiency                        &      \cmark                          &  $p_t, \alpha$                           &  \begin{tabular}[c]{@{}l@{}} Lagrange dual \\decomposition \end{tabular}                             &     \xmark                            \\ \cline{2-10} 
                               &   \begin{tabular}[c]{@{}l@{}} \textbf{This} \\\textbf{paper} \end{tabular}                         & \multicolumn{1}{l|}{$M\ge1$}  & \multicolumn{1}{l|}{$K\ge1$}  & $L\ge 1$  & Tags’ sum rate                           &  \cmark                              &  $\mathbf{w}, \mathbf{u}, \boldsymbol{\alpha}$                          &   \begin{tabular}[c]{@{}l@{}}AO \& FP \& Rayleigh\\ ratio quotient \end{tabular}                              &   \cmark                              \\ \hline
\end{tabular}
\begin{tablenotes}
 \scriptsize{\item[$\dagger$] The  variables $p_t$, $\mathbf{w}$, $\mathbf{u}$, and $\alpha$,  denote  transmit power, transmit beamforming, reception filter, and  reflection coefficient.}\\
 \scriptsize{SCA - successive convex approximation. \quad BCD - block coordinated decent.} 
 \scriptsize{\item[$\ddagger$] This work adopts NOMA to establish multiple access.}
\end{tablenotes}
\end{threeparttable}
\end{center}
\end{table*}

\subsection{Structure and Notation}
This paper is organized as follows: Section \ref{system_modelA} first introduces the system model, channel model, and EH at the tags. Next, it discusses pilot transmission and channel estimation. Further, it presents the data transmission model and the achievable rates of the tags. In Section \ref{Sec_prob_form}, we formulate the sum rate maximization problem. We present the AO solution for the proposed problem in Section \ref{Sec:Proposed_scheme}. In Section \ref{Sec_sim}, simulation examples are presented for performance evaluations. Section \ref{Sec_concl} concludes the paper and outlines future research directions.

\textit{Notation}:  
Lowercase bold and uppercase bold denote vectors and matrices.  $\mathbf{I}_n$ is  the $n \times n$ identity matrix. $\mathbf{A}^\mathrm{T}$ and $\mathbf{A}^\mathrm{H}$, denote transpose and  Hermitian transpose,  respectively.  $\mathbb{E} \{ \cdot\}$ denotes the statistical expectation. Finally, $\mathcal{CN}(\bm{\mu},\mathbf R ) $ is a complex Gaussian  vector with  mean   $\bm \mu$ and co-variance matrix  $\mathbf R$. Finally, $\mathcal{M}=\{1,\ldots,M\}$, $\mathcal{K}=\{1,\ldots,K\}$, and $\mathcal{K}_k= \mathcal{K}/k$.

 \begin{figure}[!t]\centering \vspace{-0mm}
 	\def\svgwidth{220pt} 
 	\fontsize{8}{8}\selectfont 
 	\graphicspath{{Figures/}}
 	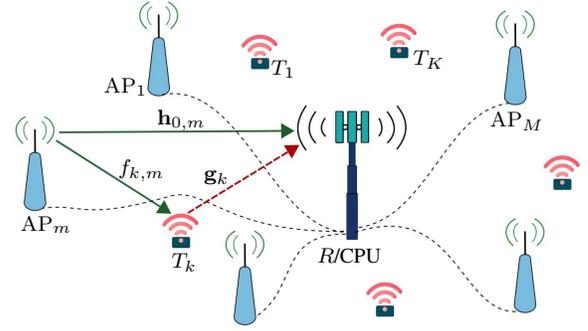 \vspace{-0mm}
 	\caption{Cell-free \bbc setup.}\vspace{-0mm} \label{fig_system_fig}
 \end{figure}

\section{System, Channel, and Signal Models}\label{system_modelA}

\subsection{System Model}
We consider a cell-free \bbc system with $M$ dedicated hybrid access point (H-AP), denoted by ${\rm{AP}}_m, m \in \mathcal{M}$, $K$ single-antenna tags, denoted by $T_k, k \in \mathcal{K}$, and a $L$-antenna reader, denoted by $R$ (Fig.~\ref{fig_system_fig}). Each H-AP equipped with a single antenna operates in either FD or HD mode \cite{Yerzhanova2021}. Specifically, in FD mode, the antenna of an H-AP simultaneously excites a carrier signal and receives signals reflected by tags. We assume the perfect FD operation with perfect self-interference cancellation at the H-AP's decoupler \cite{Liao2020, Yang2019}.  Each tag modulates its own data on the RF signals transmitted by the  H-APs and transmits the modulated signals to the reader. We assume that the reader comprises a CPU, which is connected to all H-APs via a front-haul/back-haul link \cite{Demir2021book}. Hence, this front-haul/back-haul connection assists to share the necessary CSI between the reader and the H-APs. It also assists in the synchronization between all APs to simultaneously serve all tags in the same time-frequency resource block by adopting spatial multiplexing rendered by cell-free massive MIMO.

\subsection{Channel Model}

We consider a block flat-fading channel model. During each fading block, $\mathbf{h}_{0,m} \in \mathbb{C}^{L\times 1}$, $f_{k,m} \in \mathbb{C}$, and $\mathbf{g}_k \in \mathbb{C}^{L\times 1}$  denote the channels between ${\rm{AP}}_m-R$, ${\rm{AP}}_m-T_k$, and $T_k-R$, respectively. Moreover, $\mathbf{h}_{k,m}=f_{k,m} \mathbf{g}_k \in \mathbb{C}^{L\times 1}$ denotes the cascaded channel between ${\rm{AP}}_m-T_k-R$. A unified representation of all individual channels is given as
\begin{eqnarray} \label{eqn_channel_model}
    \mathbf{v} = \zeta_v^{1/2} \tilde{\mathbf{v}},
\end{eqnarray}
where $\mathbf{v} \in \{\mathbf{h}_{0,m}, f_{k,m}, \mathbf{g}_k\}$. In \eqref{eqn_channel_model}, $\zeta_v$ captures the large-scale path-loss and shadowing,  which stays constant for several coherence intervals. Hence, the channel statistics at the H-APs and reader are assumed to be known a-prior since they change very slowly.
Moreover, $\tilde{\mathbf{v}} \sim \mathcal{CN}(\mathbf{0}, \mathbf{I}_{N})$ accounts for the small-scale Rayleigh fading\footnote{Note that $f_{k,m} = \zeta_{f_{k,m}}^{1/2} \tilde{f}_{k,m}$ and $\tilde{f}_{k,m} \sim  \mathcal{CN}\left(0,1 \right)$.}, where $N\in \{L,1\}$.

\subsection{EH at Tags} \label{sec_tag_EH}
In the considered system, tags are assumed to be semi-passive tags with small energy storage \cite{Galappaththige2023}. In particular, tags only use the stored energy during the channel estimation phase. Otherwise, tags perform EH and data transmission simultaneously via the power-splitting mode in the data transmission phase \cite{Galappaththige2023}. 

During data transmission, $T_k$ reflects $\alpha_k P_k$  for data transmission and absorbs  $P_{k}^{in}=(1-\alpha_k) P_k$, for EH, where  $P_k$  is incident RF power and  $\alpha_k \in (0, 1)$ is the reflection coefficient of $T_k$.
The EH circuit converts  $P_{k}^{in}$ to direct current (DC) power to perform internal operations and refill the energy  storage \cite{Galappaththige2022}. The amount of harvested power at $T_k$,  $P_{h,k}$, can be modeled as a linear or nonlinear function of the incident RF power. The linear model is the most widely used in the literature due to its simplicity but ignores the non-linear characteristics of actual EH circuits such as saturation and sensitivity. For that reason, several non-linear models have been developed \cite{Wang2020WPCN}.  In particular, for the nonlinear model, $P_{h,k} = \Phi(P_k^{in})$, where $\Phi(.)$ is the nonlinear EH function \cite[\textit{Eqn. (2)}]{Galappaththige2022SWIPT}, and for the linear model, $P_{h,k} = \eta_b P_k^{in}$, where $\eta_b \in (0,1]$ is the power conversion efficiency, which is typical $\eta_b=\{0.2, 0.4, 0.6\}$ \cite{Galappaththige2023}. Fortunately, both liner and non-liner models can be handled under a single unified framework, as we show next. 

In order to activate the tag,   the harvested power should exceed the threshold, i.e.,   $P_{h,k} \ge p_b$. The threshold is about \qty{-20}{\dB m} for commercial passive tags \cite{Galappaththige2022}. In particular, $P_{k}^{in} \geq  p_b'$, where $  p_b' \triangleq \Phi^{-1}(p_b) $ for nonlinear EH model and $p_b'=p_b/\eta_b$ for linear EH case.

\subsection{Pilot Transmission and Channel Estimation} \label{channel_estimate}

In our study, we adopt a cell-free \bbc system operating in the time division duplex (TDD) transmission mode for both channel estimation and data transmission \cite{Long2020}. TDD, a widely used multiple-access technique, allocates distinct time slots for uplink (UL) and downlink (DL) transmissions over a single channel. This characteristic makes TDD an efficient choice in terms of spectrum utilization \cite{tse_viswanath_2005}, as it allows dynamic allocation of time slots without the need to modify the bandwidth. This flexibility is beneficial in meeting diverse application requirements and ensuring the quality of service \cite{tse_viswanath_2005}, making TDD particularly suitable for scenarios with unpaired spectrum and asymmetric data rate needs. One of the advantages of TDD over frequency division duplex (FDD) is its simplified channel equalization methods, emanating from the principle of channel reciprocity \cite{tse_viswanath_2005}. This simplification reduces the complexity of the system while maintaining reliable communication performance. In sum, Our choice of TDD for the cell-free \bbc system is justified by its spectrum efficiency, dynamic resource allocation, and simplicity compared to FDD \cite{tse_viswanath_2005}. 


Accurate estimation of the channels is necessary to fully benefit from cell-free \bbc. In particular, the CSI of forward channels, i.e., $f_{k,m}$ for $ m \in \mathcal{M}, k \in \mathcal{K}$, and the cascaded channels, i.e., $\mathbf{h}_{k,m}$ for $ m \in \mathcal{M}, k \in \mathcal{K}$, is crucial for designing beamforming at APs, reflection coefficients at tags, and reception filters at the reader while guaranteeing EH and rate performance. On the other hand, the reader requires accurate information about the direct channels, i.e., $\mathbf{h}_{0,m} $ for $ m \in \mathcal{M}$, to suppress the direct link interference before decoding the tags' data. By accurately estimating these channels, the \bbc system's performance can be enhanced by improving signal quality and successfully decoding data transmitted by tags.

\begin{figure}[!t]\centering \vspace{0mm}
	\def\svgwidth{250pt} 
	\fontsize{8}{8}\selectfont 
	\graphicspath{{Figures/}}
 	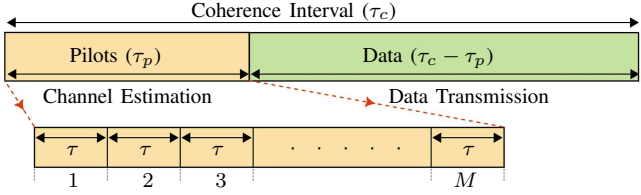 \vspace{-2mm}
	\caption{Transmission framework.}\vspace{-0mm} \label{fig:system_model}
\end{figure}
Following the methodology presented in \cite{rezaei2023timespread}, we employ a pilot-based channel estimation scheme to estimate $\mathbf{h}_{0,m}$, $\mathbf{h}_{k,m}$, and $f_{k,m}$ for $ m \in \mathcal{M}$ and $k \in \mathcal{K}$. We assume that in each coherence block of length $\tau_c$, $\tau_p$($<\tau_c$) samples are used for channel estimation (Fig. \ref{fig:system_model}). During the channel estimation phase, we adopt a TDMA scheme, using an on-off switching protocol that is controlled by the CPU.  Specifically, the channel estimation duration $\tau_p$ is divided into $M$ slots, each with length $\tau \ge K+1$, and $\tau_p = M \tau$. In each time slot, the CPU activates a specific AP in FD mode to transmit a pilot sequence $\mathbf{s} =[s_1,\ldots,s_{\tau}]\in \mathbb{C}^{1 \times \tau}$, while the remaining APs are turned off. Here, $s_i$  satisfies $|{s}_i|^2 = 1$ for $ i = \{1.\ldots, \tau\}$. In the given time slot, all the tags are active and backscatter the AP signal to transmit their pilot signals, i.e., $T_k$ backscatters $\mathbf{c}_k  = [c_{k1},\ldots c_{k \tau}]\in \mathbb{C}^{1 \times \tau},$ where  $c_{ki}$ is the tag's transmit pilot symbol over the $i$th AP symbol, $s_i$. The reader estimates the direct and cascaded channels using the tags' backscattered signals and the activated AP signal, whereas the activated AP estimates the forward channels using the tags' backscattered signals. 

\begin{rem}
    For evaluating channel estimation performance, we assume that the tags use modified Zadoff-Chu (ZC) sequences as pilots, $\mathbf{c}_k$, as in \cite{rezaei2023timespread}. However, other sequences, such as rows of Hadamard Matrix, can be used to estimate the channels. Furthermore, according to Theorem 4 in \cite{rezaei2023timespread}, any set of orthogonal sequences can be modified for backscatter channel estimation based on the characteristics of the tag's RF front end, the complexity of the tag itself, and the application environment.
\end{rem}

The received signals at the reader at the $m$th time slot,  over $\tau$ pilot symbols, can be expressed  as \cite{rezaei2023timespread}
\begin{eqnarray}\label{received_pilot}
 \mathbf{Y}_{m} = \sqrt{p_p} {\mathbf{H}_{m} } \mathbf{X} \mathbf{S} +  \mathbf{N}_{m},
\end{eqnarray}
where $p_p$ is  the pilot transmit power, $\mathbf{H}_{m} = [\mathbf{h}_{0,m}, \sqrt{\alpha_1} \mathbf{h}_{1,m}, \ldots, \sqrt{\alpha_K} \mathbf{h}_{K,m}] \in  \mathbb{C}^{L \times (K+1)}$, $\mathbf{S}\triangleq \rm{diag}(\mathbf{s})$, and $\mathbf{N}_m \in \mathbb{C}^{L \times \tau}$ is the noise matrix  with i.i.d $\mathcal{CN}(0, \sigma^2)$ elements. In \eqref{received_pilot}, $\mathbf{X} = [\mathbf{x}_1, \ldots, \mathbf{x}_{\tau}] \in \mathbb{C}^{(K+1) \times \tau}$, includes the transmitted pilots by the tags, $\mathbf{x}_i = [1, c_{1i},\ldots, c_{Ki}]^{\rm{T}} \in {\mathbb{C}^{(K+1) \times 1}}$, and $\mathbf{X}\mathbf{X}^{\rm{H}} = \tau \mathbf{I}_{K+1}$. 

In order to  estimate ${\mathbf{H}}_m $, by projecting the received signal \eqref{received_pilot} onto $\mathbf{S}^{\rm{H}}$, the post-processed received signal is given  as
\begin{eqnarray}\label{linear_final}
    \bar{\mathbf{Y}}_m= \sqrt{p_p} {\mathbf{H}}_m  \mathbf{X}+ \bar{\mathbf{N}}_m, 
\end{eqnarray}
where $\bar{\mathbf{N}}_m= \mathbf{N}_m \mathbf{S}^{\rm{H}}\in \mathbb{C}^{L \times \tau}$ is the noise matrix  with i.i.d $\mathcal{CN}(0, \sigma^2)$ elements. 

The reader then correlates  the received pilot signal in \eqref{linear_final} with $\mathbf{X}$, which results in a de-spreading operation. The post-processed signal is thus given as
\begin{eqnarray}\label{post_procss_Y}
    \bar{\mathbf{Y}}_{m,p} = \bar{\mathbf{Y}}_m \mathbf{X}^{\rm{H}}/\tau = \sqrt{p_p} {\mathbf{H}_m} + \bar{\mathbf{N}}_{m,p}, 
\end{eqnarray}
where $\bar{\mathbf{N}}_{m,p} = \bar{\mathbf{N}}_m \mathbf{X}^{\rm{H}}/\tau$ having i.i.d $\mathcal{CN}(0, \sigma_p^2)$ elements, where $\sigma_p^2=\sigma^2/{\tau}$. Given  independent Rayleigh fading, the elements of the channel matrix and the noise matrix are statistically independent. Next, the $(l,k)$th element of \eqref{post_procss_Y} is given as
\begin{eqnarray}\label{post_procss_mk}
    [\bar{\mathbf{Y}}_{m,p}]_{l,k} \triangleq y_{l,k}^m = \sqrt{p_p} {h}_{l,k}^m + n_{l,k}^m, 
\end{eqnarray}
where ${h}_{l,k}^m =[{\mathbf{H}}_m]_{l,k}$ and $n_{l,k}^m = [\bar{\mathbf{N}}_{m,p}]_{l,k}$.
 
The MMSE estimator is thus given as \cite{ rezaei2023timespread}
\begin{eqnarray}\label{mmse_h}
    \hat{{h}}_{l,k}^m &=& \mathbb{E}\{{h}_{l,k}^m \vert  y_{l,k}^m \} = \frac{\mathbb{E}\{\bar{h}_{l,k}^m  (y_{l,k}^m)^* \}}{ \mathbb{E}\{ \vert  y_{l,k}^m \vert^2 \}} y_{lk}^m \nonumber \\
    &=&  \begin{cases}
     \frac{\sqrt{p_p} \zeta_{\mathbf{h}_{0,m}}}{ p_p \zeta_{\mathbf{h}_{0,m}} + \sigma_p^2} y_{l0}^m, \quad \text{for} \quad k=0, \\
     \frac{\sqrt{\alpha_k p_p} \zeta_{\mathbf{h}_{k,m}}}{\alpha_k p_p  \zeta_{\mathbf{h}_{k,m}} + \sigma_p^2} y_{lk}^m, \quad \text{for} \quad k\in \mathcal{K}, \\
    \end{cases}
\end{eqnarray}
where $\zeta_{\mathbf{h}_{k,m}} = \zeta_{{f}_{k,m}} \zeta_{\mathbf{g}_{k}}$.

Thus, the MMSE estimate of the complete channel matrix, ${\mathbf{H}}_m$,  is given as
\begin{eqnarray}\label{MMSE}
    \hat{{\mathbf{H}}}_m^{\rm{MMSE}} =  \mathbf{Y}_{m,p} \mathbf{D}_{m,\gamma}^{1/2},
\end{eqnarray}
where $\mathbf{D}_{m,\gamma} = {\rm{diag}}([\gamma_{0,m}, \gamma_{1,m}, \ldots, \gamma_{K,m} ])$, in which $\gamma_{0,m} = \frac{ p_p \zeta_{\mathbf{h}_{0,m}}^2}{ p_p \zeta_{\mathbf{h}_{0,m}} + \sigma_p^2}$ and $\gamma_{k,m} = \frac{\alpha_k p_p \zeta_{\mathbf{h}_{k,m}}^2}{\alpha_k p_p \zeta_{\mathbf{h}_{k,m}} + \sigma_p^2}.$

Additionally, using \eqref{received_pilot}, the LS estimate which operates without any prior knowledge of the channel can be obtained as
\begin{eqnarray}\label{LS_estimate}
         \hat{{\mathbf{H}}}_m^{\rm{LS}} =  \mathbf{Y}_m \bar{\mathbf{X}}^{\dagger}, 
        \end{eqnarray}
where $\bar{\mathbf{X}}^{\dagger} = {{\mathbf{X}}}^{\rm{H}}(\bar{\mathbf{X}}{\bar{\mathbf{X}}}^{\rm{H}})^{-1}$, where $\bar{\mathbf{X}} = \sqrt{p_p}{\mathbf{X}}$.
 
Similarly, the received signal at  ${\rm{AP}}_{m}$  over $\tau$ pilot symbols can be  expressed  as 
\begin{eqnarray}\label{received_pilot_AP_1}
    \mathbf{y}_{m}^p = \sqrt{p_p} \sum_{i\in \mathcal{K}} \sqrt{\alpha_i}{f}_{i,m}^{\rm{T}} {f}_{i,m} \mathbf{c}_i \mathbf{S} + \mathbf{n}_m.
\end{eqnarray}
where $\mathbf{n}_m \in \mathbb{C}^{1\times \tau} \sim \mathcal{CN}(\mathbf{0},\sigma_a^2 \mathbf{I}_{\tau})$ denotes the noise at ${\rm{AP}}_m$. 
The post-processed received signal is thus given as 
\begin{eqnarray}\label{received_pilot_AP_2}
    \bar{\mathbf{y}}_{m}^p = \mathbf{y}_{m}^p \mathbf{S}^{\rm{H}} = \sqrt{p_p} \sum_{i\in \mathcal{K}} \sqrt{\alpha_i}{f}_{i,m}^{\rm{T}} {f}_{i,m} \mathbf{c}_i + \bar{\mathbf{n}}_m,
\end{eqnarray}
where $\bar{\mathbf{n}}_m = \mathbf{n}_m \mathbf{S}^{\rm{H}}$. The received pilot signal is then projected onto the $\mathbf{c}_k^{\rm{H}}$ which yields
\begin{eqnarray}\label{received_pilot_AP_3}
    y_{k,m}^p = \bar{\mathbf{y}}_{m}^p \mathbf{c}_k^{\rm{H}}/\tau = \sqrt{p_p\alpha_k} {f}_{k,m}^{\rm{T}} {f}_{k,m} + {n}_{k,m},
\end{eqnarray}
where  ${n}_{k,m} = \bar{\mathbf{n}}_{m}\mathbf{c}_i^{\rm{H}}/\tau  \sim \mathcal{CN}(0,\sigma_a^2/\tau )$.

Therefore, the LS estimate of $\bar{f}_{k,m} = {f}_{k,m}^{\rm{T}} {f}_{k,m}$ can be obtained as\footnote{For the forward link we only derive the LS estimator for brevity.}
\begin{eqnarray}\label{LS_estimate_f}
     \hat{\bar{f}}_{k,m}^{\rm{LS}} =  y_{k,m}^p/\sqrt{p_p\alpha_k}.
\end{eqnarray}
Hence, we can obtain an approximate estimate for the forward channel, $f_{k,m}$, using the estimates of $\bar{f}_{k,m}$, as $\hat{f}_{k,m}^{\rm{LS}} \approx \sqrt{\hat{\bar{f}}_{k,m}^{\rm{LS}}}$.

\subsection{Transmission Model} 
All APs simultaneously serve the tags in HD mode in the data transmission phase. The signal transmitted at ${\rm{AP}}_m$ is thus given as
\begin{eqnarray}
    q_m = \sqrt{p_t} \sum_{i \in \mathcal{K}} w_{i,m} s,
\end{eqnarray}
where $p_t$ is the transmit power at each AP, $w_{i,m}\in  \mathbb{C}$ is the spatial directivity/precoder of the signal at ${\rm{AP}}_m$ for $T_i$, and $s \sim \mathcal{CN}(0,1)$ is the carrier signal satisfying $\mathbb{E}\{|s|^2\}=1$. The signal received at $T_k$ is given as
\begin{eqnarray} \label{eqn_rx_tag}
    y_k = \mathbf{f}_k^{\rm{T}} \mathbf{q}  = \sqrt{p_t} \sum_{i\in \mathcal{K}} \mathbf{f}_k^{\rm{T}} \mathbf{w}_i s,
\end{eqnarray}
where $\mathbf{q}=[q_1,\ldots,q_M]^{\rm{T}} \in \mathbb{C}^{M \times 1}$,  and $\mathbf{w}_i=[w_{1i}, \ldots,w_{Mi}]^{\rm{T}}\in \mathbb{C}^{M \times 1}$.  Moreover, $\mathbf{f}_k = [f_{k,1}, \ldots, f_{k,M}]^{\rm{T}} \in \mathbb{C}^{M \times 1}$ denotes the effective channel between all APs  and $T_k$.

The tags must harvest enough power to support their internal operations. Hence, the input signal power at EH circuit of  $T_k$ must satisfy the following energy constraint (Section \ref{sec_tag_EH}):
\begin{eqnarray} \label{eqn_EH_tag}
  P_{k}^{in} =   (1-\alpha_k) p_t \left\vert \sum\nolimits_{i\in \mathcal{K}} \mathbf{f}_k^{\rm{T}} \mathbf{w}_i \right\vert^2 \geq p_b',
\end{eqnarray}
where $  p_b' \triangleq \Phi^{-1}(p_b) $ for nonlinear EH model and $p_b'=p_b/\eta_b$ for linear EH case. 
$T_k$ harvests energy from the received signal and modulates it with its data, $c_k$, where $c_k$ is the normalized backscatter symbol selected from a multi-level ($\bar{M}$-ary) modulation such that $\mathbb{E}\{|c_k|^2\}=1$,  before sending it to the reader. The received signal at the reader is thus given as
\begin{eqnarray} \label{eqn_rx_reader}
    \mathbf{r} &=& \sum_{m\in \mathcal{M}} \mathbf{h}_{0,m} q_m + \sum_{j \in \mathcal{K}} \sqrt{\alpha_j} \mathbf{g}_j  y_j c_j + \mathbf{z} \nonumber \\
    &=& \mathbf{H}_0 \mathbf{q} +  \sqrt{p_t} \sum_{j\in \mathcal{K}} \sum_{i\in \mathcal{K}} \sqrt{\alpha_j}  \mathbf{g}_j \mathbf{f}_j^{\rm{T}} \mathbf{w}_i s c_j + \mathbf{z},
\end{eqnarray}
where $\mathbf{H}_0=[\mathbf{h}_{0,1}, \ldots \mathbf{h}_{0,M}] \in \mathbb{C}^{L \times M}$,  and 
$\mathbf{z}\in \mathbb{C}^{L \times 1} \sim \mathcal{CN}(\mathbf{0}, \sigma^2\mathbf{I}_{L})$ is the additive white Gaussian noise (AWGN) vector at the reader. In \eqref{eqn_rx_reader}, the first term is the direct-link signals from the APs and the second term is the backscatter-link signals from tags. 

\begin{rem}\label{rem_direct_channel}
    The reader uses SIC to remove the direct-link signals from all APs and then applies the combining filter, $\mathbf{u}_k \in \mathbb{C}^{L \times 1}$ to decode $T_k$'s signal.  Because direct channels between APs and the reader are conventional cell-free one-way channels, they can be estimated more accurately, i.e. with a lower normalized mean square error than backscatter channels.  Our simulation results also support the validity of this argument (Fig~\ref{fig_NMSE_Ptx}).
\end{rem}

Following Remark~\ref{rem_direct_channel}, we assume that the reader is able to perfectly remove the direct APs' signals from the received signal \eqref{eqn_rx_reader}. The post-processed signal for decoding $T_k$'s data at the reader is thus given as
\begin{eqnarray} \label{eqn_rx_reader_k}
    r_k &=& \mathbf{u}_k^{\rm{T}} \left(\mathbf{r} -\mathbf{H}_0 \mathbf{q} \right) \nonumber \\
    &=& \underbrace{\sqrt{\alpha_k p_t} \sum_{i\in \mathcal{K}} \mathbf{u}_k^{\rm{T}} \mathbf{g}_k   \mathbf{f}_k^{\rm{T}} \mathbf{w}_i s c_k }_{\text{Desired signal}} \nonumber \\
    &&+ \underbrace{\sqrt{ p_t} \sum_{j\in \mathcal{K}_k}   \sum_{i\in \mathcal{K}} \sqrt{\alpha_j} \mathbf{u}_k^{\rm{T}} \mathbf{g}_j   \mathbf{f}_j^{\rm{T}} \mathbf{w}_i s c_j }_{\text{Interference from the tag}} + \underbrace{\mathbf{u}_k^{\rm{T}} \mathbf{z}}_{\text{Noise}}, \quad
\end{eqnarray}
where the first, second, and third terms in $r_k$ are respectively the desired signal, interference from other tags, and effective noise at the reader. 

\subsection{Achievable Rate} 
This section derives the achievable rates of the tags. Using \eqref{eqn_rx_reader_k}, the received SINR for $T_k$, $\gamma_k'$, at the reader is obtained as
\begin{eqnarray}\label{eqn_SINR_tag}
    \gamma_k' = \frac{\alpha_k p_t \left\vert    \sum_{i\in \mathcal{K}} \mathbf{u}_k^{\rm{T}} \mathbf{g}_k \mathbf{f}_k^{\rm{T}} \mathbf{w}_i     \right\vert^2 \vert s \vert^2}{ p_t  \sum_{j\in \mathcal{K}_k} \alpha_j \left\vert    \sum_{i\in \mathcal{K}}  \mathbf{u}_k^{\rm{T}} \mathbf{g}_j \mathbf{f}_j^{\rm{T}}  \mathbf{w}_i  \right\vert^2 \vert s \vert^2+  \Vert \mathbf{u}_k  \Vert^2   \sigma^2}. ~~
\end{eqnarray} 
Thus, the achievable rate of $T_k$ at the reader is given as
\begin{eqnarray}\label{eqn_rate_tag}
	\mathcal{R}_k = \psi \mathbb{E}_s \{ {\rm{log}}_2(1+\gamma_k') \},
\end{eqnarray} 
where the pre-log factor $\psi=(\tau_c-\tau_p)/\tau_c$ captures the effective portion of the coherence interval for the DL transmission. 
By taking the average over $s$ in \eqref{eqn_rate_tag}, the rate of $T_k$ is computed as\footnote{For  $s \sim \mathcal{CN}(0,1)$, $|s|^2$ is exponentially  distributed.}
\begin{eqnarray}\label{eqn:rate_Tk_true}
     \!\!\! \mathcal{R}_{k} \!= \psi {\log}_2(e) \! \left(\!-e^{\frac{1}{a_{k}+b_{k}}} {\mathrm{E}}_i\!\left(\frac{-1}{a_{k}+b_{k}}\right) \! + e^{\frac{1}{b_{k}}} {\mathrm{E}}_i\!\left(\frac{-1}{b_{k}}\right)\! \right)\!,\,\,\,
\end{eqnarray}
where 
\begin{subequations}\label{a_k_b_k_def}
    \begin{eqnarray}
a_{k} &\triangleq& \frac{\alpha_k p_t}{\Vert \mathbf{u}_k  \Vert^2 \sigma^2 }  \left\vert    \sum\nolimits_{i\in \mathcal{K}} \mathbf{u}_k^{\rm{T}} \mathbf{g}_k \mathbf{f}_k^{\rm{T}} \mathbf{w}_i     \right\vert^2, \\
b_k &\triangleq& \frac{p_t}{ \Vert \mathbf{u}_k \Vert^2 \sigma^2}  \sum\nolimits_{j\in \mathcal{K}_k} \alpha_j \left\vert    \sum\nolimits_{i\in \mathcal{K}}  \mathbf{u}_k^{\rm{T}} \mathbf{g}_j \mathbf{f}_j^{\rm{T}}  \mathbf{w}_i  \right\vert^2 .  \qquad 
\end{eqnarray}
\end{subequations}
Additionally, ${\mathrm{E}}_i(x) = \int_{-\infty}^{x} u^{-1} e^u du$ is the exponential integral function.  Note  that, $-e^{\frac{1}{x}} {\mathrm{E}}_i ({-1}/{x})$ is monotonically increasing and concave function of $x$ \cite{Long2020}. 

\section{Optimization Problem Formulation}\label{Sec_prob_form}
Herein, we optimize the performance of the proposed cell-free \bbc setup (Fig.~\ref{fig_system_fig}). We aim to maximize the sum rate of the tags while ensuring the tag activation, i.e., the tags meet the minimum operational energy requirements. The optimization variables are APs' precoders, i.e., $\mathbf{w}_k$ for $k \in \mathcal{K}$, the combining filters at the reader, i.e,  $\mathbf{u}_k$ for $k \in \mathcal{K}$, and the tags' reflection coefficients, i.e, $\alpha_k$ for $k \in \mathcal{K}$. 

The sum rate maximization problem can be formulated as
\begin{subequations}\label{P1_prob}
    \begin{eqnarray}
        \mathbf{P1}:
        \underset {\mathbf{w}_k, \mathbf{u}_k, \alpha_k, \forall k}{\text{maximize}} && \sum\nolimits_{k\in \mathcal{K}}  \mathcal{R}_k, \label{P1_obj} \\
        \text{subject to} && \sum\nolimits_{i\in \mathcal{K}} \vert w_{i,m} \vert^2 \leq 1, \label{P1_AP_power} \\
        && (1-\alpha_k) p_t \left\vert \sum\nolimits_{i\in \mathcal{K}} \mathbf{f}_k^{\rm{T}} \mathbf{w}_i \right\vert^2 \geq p_b', \quad \label{P1_tag_EH}\\
        &&\Vert \mathbf{u}_k\Vert^2 \leq 1, \label{P1_detector}\\
        &&0 < \alpha_k < 1. \label{alpha_constr}
    \end{eqnarray}
\end{subequations}
Here, \eqref{P1_AP_power}  is the per-AP transmit power constraint, \eqref{P1_tag_EH}  is the minimum per-tag power requirement for EH, and \eqref{P1_detector} is the normalization constraint for the combining filter at the reader.

\begin{rem}
    In order to solve the proposed sum rate maximization problem, we replace the rate of each tag, given in \eqref{eqn_rate_tag}, with the following bound for traceability \cite{Zhang2014}:
    \begin{eqnarray}
        \mathcal{R}_k  \approx \psi{\rm{log}}_2(1+ \mathbb{E}_s \{\gamma_k'\}) = \psi {\rm{log}}_2(1+  \gamma_k),
    \end{eqnarray}
    where 
    \begin{eqnarray}\label{eqn_SINR_tag_bound}
        \gamma_k = \frac{\alpha_k p_t \left\vert    \sum_{i\in \mathcal{K}} \mathbf{u}_k^{\rm{T}} \mathbf{g}_k \mathbf{f}_k^{\rm{T}} \mathbf{w}_i     \right\vert^2 }{ p_t  \sum_{j\in \mathcal{K}_k} \alpha_j \left\vert    \sum_{i\in \mathcal{K}}  \mathbf{u}_k^{\rm{T}} \mathbf{g}_j \mathbf{f}_j^{\rm{T}}  \mathbf{w}_i  \right\vert^2 +  \Vert \mathbf{u}_k  \Vert^2   \sigma^2}. ~~
    \end{eqnarray} 

   In addition, since the CPU/reader performs the optimization by using the estimated CSI of the channels, all of the objective functions and constraints in $\mathbf{P1}$ are substituted with their respective channel estimations.
\end{rem}

The proposed sum rate maximization, $\mathbf{P1}$, has an objective function and constraints that are not convex functions in $\mathbf{w}_k$, $\mathbf{u}_k$, and $\alpha_k$. Non-convex problems are hard to solve optimally. Such a problem may have multiple feasible regions and multiple locally optimal points within each region. The problem of
finding the exact global solution of a non-convex problem is NP-hard.  Therefore,  we resort to decoupling the optimization variables and employ AO, FP, and Rayleigh ratio quotient approaches.

\section{Proposed Solution}\label{Sec:Proposed_scheme}
As mentioned before, the AO paradigm requires the optimization variables to be split into non-overlapping blocks of one or more block variables. This allows iterative optimization of  one block variable at a time while holding the others fixed, then proceeds on to optimizing the next block while holding the others fixed, and so on until convergence is attained. This method is also known as block coordinate descent \cite{Stanczak2009book}.  In our problem, $\mathbf{w}_k$, $\mathbf{u}_k$, and $\alpha_k$ are the natural choice of blocks.

\subsection{Transmit Beamforming}
When the tags' reflection coefficients and the reader's receiver combining are fixed, $\mathbf{P1}$ reduces to the following transmit beamforming optimization problem: 
\begin{subequations}\label{Pw_prob}
    \begin{eqnarray}
        \mathbf{P}_{\mathrm{w}}:
        \underset {\mathbf{w}_k \forall k}{\text{maximize}} && \sum\nolimits_{k\in \mathcal{K}}  \psi {\rm{log}}_2(1+  \hat{\gamma}_k), \label{Pw_obj} \\
        \text{subject to} && \sum\nolimits_{i\in \mathcal{K}} \vert w_{i,m} \vert^2 \leq 1, \label{Pw_AP_power} \\
        && (1-\alpha_k) p_t \left\vert \sum\nolimits_{i\in \mathcal{K}} \hat{\mathbf{f}}_k^{\rm{T}} \mathbf{w}_i \right\vert^2 \geq p_b', \quad \label{Pw_tag_EH}
    \end{eqnarray}
\end{subequations}
where $\hat{\gamma}_k$ is obtained by replacing the respective channels, i.e., $\mathbf{g}_k$ and $\mathbf{f}_k$ for $k \in \mathcal{K}$, in \eqref{eqn_SINR_tag_bound} with respective estimated channels. 

Next, we define $\mathbf{a}_{kj} \triangleq [\mathbf{u}_k^{\rm{T}} \hat{\mathbf{g}}_j \hat{\mathbf{f}}_j^{\rm{T}}, \ldots, \mathbf{u}_k^{\rm{T}} \hat{\mathbf{g}}_j \hat{\mathbf{f}}_j^{\rm{T}}] \in \mathbb{C}^{MK \times 1}$ and $\mathbf{w} \triangleq [\mathbf{w}_1^{\rm{T}}, \ldots, \mathbf{w}_K^{\rm{T}}]^{\rm{T}} \in \mathbb{C}^{MK \times 1}$. Thereby, the $T_k$'s SINR in \eqref{eqn_SINR_tag_bound} is rearranged as
\begin{eqnarray}\label{eqn_SINR_tag_bound1}
    \hat{\gamma}_k = \frac{\alpha_k p_t \left\vert    \mathbf{a}_{kk}^{\rm{T}} \mathbf{w}     \right\vert^2 }{ p_t  \sum_{j\in \mathcal{K}_k} \alpha_j \left\vert \mathbf{a}_{kj}^{\rm{T}} \mathbf{w}  \right\vert^2 +    \sigma_w^2},
\end{eqnarray} 
where $\sigma_w^2=\Vert \mathbf{u}_k  \Vert^2   \sigma^2$. Then, $\mathbf{P}_{\mathrm{w}}$ can be treated as a multiple ratio FP problem \cite{Shen2018, Galappaththige2022Weighted}.  We next apply a quadratic transform to the objective function of
$\mathbf{P}_{\mathrm{w}}$ as
\begin{eqnarray} \label{eqn_Pw_obj_re}
    f(\mathbf{w}, \boldsymbol{\lambda})  &=&  \sum\nolimits_{k \in \mathcal{K}} \psi {\rm{log}}_2 \left(1+ 2 \lambda_k \sqrt{\alpha_k p_t} {\rm{Re}}\left\{ \mathbf{a}_{kk}^{\rm{T}} \mathbf{w} \right\} \right. \nonumber \\
	&& \left. -\lambda_k^2 \left(p_t \sum\nolimits_{j\in \mathcal{K}_k} \alpha_j \vert \mathbf{a}_{kj}^{\rm{T}} \mathbf{w} \vert^2 + \sigma_{w}^2  \right)\right),
\end{eqnarray}
where $\boldsymbol{\lambda}=[\lambda_1, \ldots, \lambda_K]^{\rm{T}}$ is the auxiliary variable introduced by the quadratic transformation. Thereby, we alternatively optimize $\mathbf{w}$ and $\boldsymbol{\lambda}$. For a given $\mathbf{w}$, the optimal $\boldsymbol{\lambda}$ is found in closed-form as \cite{Shen2018}
\begin{eqnarray}\label{eqn_opt_lambda}
	\lambda_k^o = \frac{ \sqrt{\alpha_k p_t} {\rm{Re}} \left\{ \mathbf{a}_{kk}^{\rm{T}} \mathbf{w} \right\}}{ \ln(2)\left( p_t\sum_{j\in \mathcal{K}_k} \alpha_j \vert \mathbf{a}_{kj}^{\rm{T}} \mathbf{w} \vert^2 + \sigma_{w}^2\right)}.
\end{eqnarray}

\begin{rem}
Without losing generality, we constrain the transmit beamforming vector, $\mathbf{w}$, with the channel responses to obtain a non-negative real desired signal term, i.e., $\left| \mathbf{a}_{kk}^{\rm{T}} \mathbf{w}\right| \approx {\rm{Re}} \{\mathbf{a}_{kk}^{\rm{T}} \mathbf{w}\}$. Our simulation findings support the validity of this technique. This is due to the fact that our method iteratively maximizes the achievable rate/SINR by co-phasing the desired signal component while reducing interference.
\end{rem}

Next, we must optimize $\mathbf{w}$ for a given $\boldsymbol{\lambda}$. First, by applying several mathematical manipulations, the objective function in \eqref{eqn_Pw_obj_re} can be rearranged as
\begin{eqnarray}\label{eqn_Pw_obj_re1}
	\! f(\mathbf{w}) = \!\sum_{k\in \mathcal{K}} \psi{\rm{log}}_2 \left( 1 - \mathbf{w}^{\rm{T}}  \mathbf{U}_k \mathbf{w} + 2 {\rm{Re}} \left\{ \mathbf{w}^{\rm{T}} \mathbf{v}_k \right\} + t_k \right)\!,\,\,\,
\end{eqnarray}
where $\mathbf{U}_k$, $\mathbf{v}_k$, and $t_k$ for $k \in \mathcal{K}$ are defined as
\begin{subequations} \label{eqn_def_Uvq}
\begin{eqnarray}
    \mathbf{U}_k &\triangleq& (\lambda_k^o)^2 p_t \sum_{j\in \mathcal{K}_k} \alpha_j   \mathbf{a}_{kk} \mathbf{a}_{kk}^{\rm{T}}, \\
    \mathbf{v}_k &\triangleq& 2 \lambda_k^o \sqrt{\alpha_k p_t} \mathbf{a}_{kk},\\
    t_k &\triangleq& (\lambda_k^o)^2 \sigma_{w}^2. 
\end{eqnarray}
\end{subequations}
Next, for a given $\boldsymbol{\lambda}$, the corresponding optimization problem is given as
\begin{subequations}\label{Pw1_prob}
    \begin{eqnarray}
        \mathbf{P}_{\mathrm{w}1}:
        \underset {\mathbf{w} }{\text{maximize}} && f(\mathbf{w}), \label{Pw1_obj} \\
        \text{subject to} && \sum\nolimits_{i\in \mathcal{K}} \vert w_{i,m} \vert^2 \leq 1, \label{Pw1_AP_power} \\
        && (1-\alpha_k)p_t P_{k}^{\rm{Lin}} \geq p_b', \quad \label{Pw1_tag_EH}
    \end{eqnarray}
\end{subequations}
where $P_{k}^{\rm{Lin}}$ is the linearized received signal power at $T_k$, given as
\begin{eqnarray}\label{eqn:PT_lin}
     P_{k}^{\rm{Lin}} &=&  \left(\mathbf{w}^{(i-1)}\right)^{\rm{T}} \mathbf{D}_k \mathbf{w}^{(i-1)} \nonumber \\
    &&+ \left(\left(\mathbf{D}_k + \mathbf{D}_k^{\rm{T}} \right) \mathbf{w}^{(i-1)} \right)^{\rm{T}} \left(\mathbf{w}^{(i)}- \mathbf{w}^{(i-1)} \right), \quad 
\end{eqnarray}
where $\mathbf{D}_k = \mathbf{d}_k \mathbf{d}_k^{\rm{T}}$ and $\mathbf{d}_k \triangleq [\hat{\mathbf{f}}_k^{\rm{T}},\ldots, \hat{\mathbf{f}}_k^{\rm{T}}]^{\rm{T}} \in \mathbb{C}^{MK \times  1}$. Besides, $\mathbf{w}^{(i)}$ and $\mathbf{w}^{(i-1)}$ are the current and the previous iteration values of $\mathbf{w}$.

Because  $\mathbf{a}_{kj} \mathbf{a}_{kj}^{\rm{T}}$ is a positive-definite matrix, $\mathbf{U}_k$ is also a positive-definite matrix. Hence, the objective function, $f(\mathbf{w})$, is a quadratic concave function of $\mathbf{w}$. Consequently, $\mathbf{P}_{\mathrm{w}1}$ can be solved as a quadratically constrained quadratic program (QCQP) \cite{BoydConvex2004}. Algorithm~\ref{Algow} gives the solution of  $\mathbf{P}_{\mathrm{w}1}$. 

\begin{algorithm}[hbt!]
\caption{: Transmit beamforming.}\label{Algow}
\begin{algorithmic}
 \renewcommand{\algorithmicrequire}{\textbf{Initialization:}}
 \renewcommand{\algorithmicensure}{\textbf{Repeat}}
 \Require Initialize $\mathbf{w}$ to a feasible value.
\Ensure
\State \textbf{Step 1}: Update $\lambda$ by \eqref{eqn_opt_lambda}.
\State \textbf{Step 2}: Update $\mathbf{w}$ by solving $\mathbf{P}_{\mathrm{w}1}$ in \eqref{Pw1_prob}.
\end{algorithmic}
\textbf{Until} the value of the objective function converges.\\
\textbf{Output:} The optimal transmit beamforming vector $\mathbf{w}^o$. 
\end{algorithm}

\begin{rem}
The proposed optimization strategy for solving $\mathbf{w}$ once the original problem, $\mathbf{P}_{\mathrm{w}}$, is transformed into a convex problem is shown in Algorithm \ref{Algow}. An alternating optimization strategy is used to solve $\mathbf{P}_{\mathrm{w}}$ iteratively. We begin by quantifying the SINR in \eqref{eqn_SINR_tag_bound} upon initializing $\mathbf{w}$, and then we update a better solution for $\mathbf{w}$ in each iteration. This process is repeated until the normalized objective function increases less than $\epsilon=10^{-4}$.
\end{rem}

\subsection{Receiver Combining}\label{Sec:receiver_combine}
Here, we design the reader's combining filters. To this end, for fixed transmit beamforming and tags' reflection coefficients,  $\mathbf{P1}$ is reduced to a receiver combining optimization problem and is formulated as
\begin{subequations}\label{Pu_prob}
    \begin{eqnarray}
        \mathbf{P}_{\mathrm{u}}:
        \underset { \mathbf{u}_k, \forall k}{\text{maximize}} && \sum\nolimits_{k\in \mathcal{K}} \psi {\rm{log}}_2(1+  \hat{\gamma}_k), \label{Pu_obj} \\
        \text{subject to} && \Vert \mathbf{u}_k\Vert^2 \leq 1. \label{Pu_detector}
    \end{eqnarray}
\end{subequations}

We first define $\mathbf{b}_{k} \triangleq \mathbf{g}_k \sum_{i \in \mathcal{K}} \mathbf{f}_k^{\rm{T}} \mathbf{w}_i$ and rewrite the $T_k$'s SINR in \eqref{eqn_SINR_tag_bound} as
\begin{eqnarray}\label{eqn_SINR_tag_boundu}
    \hat{\gamma}_k &=& \frac{\alpha_k p_t \left\vert    \mathbf{b}_{k}^{\rm{T}} \mathbf{u}_k     \right\vert^2 }{ p_t  \sum_{j\in \mathcal{K}_k} \alpha_j \left\vert \mathbf{b}_{j}^{\rm{T}} \mathbf{u}_k  \right\vert^2 + \Vert \mathbf{u}_k \Vert^2  \sigma^2} \nonumber \\
    &=& \frac{ \mathbf{u}_k^{\rm{T}}  \mathbf{B}_{k} \mathbf{u}_k  }{ \mathbf{u}_k^{\rm{T}} \left( \sum_{j\in \mathcal{K}_k} \mathbf{B}_{j} + \sigma^2 \mathbf{I}_{L} \right) \mathbf{u}_k},
\end{eqnarray} 
where $\mathbf{B}_{j}= \alpha_j p_t \mathbf{b}_j \mathbf{b}_j^{\rm{T}}$. Since the objective function in \eqref{Pu_obj} is a non-decreasing function of its argument, it can be replaced by the SINR of $T_k$ \eqref{eqn_SINR_tag_boundu}, i.e., $\sum\nolimits_{k\in \mathcal{K}}  {\rm{log}}_2(1+  \hat{\gamma}_k) = \sum\nolimits_{k\in \mathcal{K}} \hat{\gamma}_k$. $\mathbf{P}_{\mathrm{u}}$ thus becomes a generalized Rayleigh ratio quotient problem \cite{Stanczak2008book, Wan2016}. Hence, the optimal combiner vector for $T_k$ is given as \cite{Stanczak2008book, Wan2016} 
\begin{eqnarray}\label{eqn_optml_u}
    \mathbf{u}_k^o = v_{\rm{max}}\left[\left( \sum\nolimits_{j\in \mathcal{K}_k} \mathbf{B}_{j} + \sigma^2 \mathbf{I}_{L} \right)^{-1} \mathbf{B}_{k} \right],
\end{eqnarray}
where $v_{\rm{max}}[\cdot]$ is the dominant eigenvector of the matrix \cite{Stanczak2008book, Wan2016}.  

\subsection{Reflection Coefficient Optimization}
As the final block of the proposed AO framework, here, we optimize the tags' reflection coefficients. For given transmit beamforming and receiver combining, $\mathbf{P1}$ becomes an optimization problem for tags’ reflection coefficients. The corresponding optimization problem is given as
\begin{subequations}\label{Pa_prob}
    \begin{eqnarray}
        \mathbf{P}_{\alpha}:
        \underset {\alpha_k, \forall k}{\text{maximize}} && \sum\nolimits_{k\in \mathcal{K}}  \psi{\rm{log}}_2(1+  \hat{\gamma}_k), \label{Pa_obj} \\
        \text{subject to} && (1-\alpha_k) p_t \left\vert \sum\nolimits_{i\in \mathcal{K}} \mathbf{f}_k^{\rm{T}} \mathbf{w}_i \right\vert^2 \geq p_b', \quad \label{Pa_tag_EH}\\
        &&0 < \alpha_k < 1. \label{Pa_alpha_constr}
    \end{eqnarray}
\end{subequations}
To solve $\mathbf{P}_{\alpha}$, we first introduce $\theta_k$ to replace the SINR terms in \eqref{Pa_obj}, such that $\theta_k\leq \hat{\gamma}_k$, and then reformulate $\mathbf{P}_{\alpha}$ as follows:
\begin{subequations}\label{Pa1_prob}
    \begin{eqnarray}
        \mathbf{P}_{\alpha 1}:
        \underset {\boldsymbol{\alpha}, \boldsymbol{\theta}}{\text{maximize}} && \sum\nolimits_{k\in \mathcal{K}}  \psi{\rm{log}}_2(1+  \theta_k), \label{Pa1_obj} \\
        \text{subject to} && \theta_k \leq \frac{\hat{A}_k(\boldsymbol{\alpha})}{\hat{B}_k(\boldsymbol{\alpha})},  \label{Pa1_rate}\\ 
        && (1-\alpha_k) p_t \left\vert  \mathbf{d}_k^{\rm{T}} \mathbf{w} \right\vert^2 \geq p_b', \quad \label{Pa1_tag_EH}\\
        &&0 < \alpha_k < 1, \label{Pa1_alpha_constr}
    \end{eqnarray}
\end{subequations}
where $\boldsymbol{\alpha}= [\alpha_1, \ldots, \alpha_K]^{\rm{T}}$ and $\boldsymbol{\theta}=[\theta_1,\ldots,\theta_K]^{\rm{T}}$. Besides, $\hat{A}_k(x)$ and $\hat{B}_k(x)$ are the numerators and denominators of the corresponding SINR terms  $\hat{\gamma}_k$ as functions of the variable $x$ i.e., $\boldsymbol{\alpha}$. The reformulated optimization problem $\mathbf{P}_{\alpha 1}$ can be considered as a two-part optimization problem, i.e., (i) an outer optimization over $\boldsymbol{\alpha}$ with fixed $\boldsymbol{\theta}$ and (ii) an inner optimization over $\boldsymbol{\theta}$ with fixed $\boldsymbol{\alpha}$.

The inner optimization problem is thus given as
\begin{subequations}\label{Pa2_prob}
    \begin{eqnarray}
        \mathbf{P}_{\alpha 2}:
        \underset {\boldsymbol{\alpha}, \boldsymbol{\theta}}{\text{maximize}} && \sum\nolimits_{k\in \mathcal{K}}  \psi {\rm{log}}_2(1+  \theta_k), \label{Pa2_obj} \\
        \text{subject to} && \theta_k \leq \frac{\hat{A}_k(\bm{\alpha})}{\hat{B}_k(\bm{\alpha})}.  \label{Pa2_rate} 
    \end{eqnarray}
\end{subequations}
This inner optimization problem in \eqref{Pa2_prob} is convex in $\boldsymbol{\theta}$ and has strong duality \cite{Shen2018}. Thus, he solution to $\mathbf{P}_{\alpha 2}$ is that $\theta_k$ satisfies \eqref{Pa2_rate} with equality. We use the Lagrangian dual transform \cite{Shen2018} to deal with the logarithm in the objective function of $\mathbf{P}_{\alpha 2}$, and the corresponding Lagrangian function is given as
\begin{eqnarray} \label{LD_func_a}
L(\boldsymbol{\theta},\boldsymbol{\mu})= \!\sum_{k\in \mathcal{K}} \psi{\log}_2(1+\theta_{k}) -\sum_{k\in \mathcal{K}} \mu_k \left(\theta_k -\frac{\hat{A}_k(\boldsymbol{\alpha})}{\hat{B}_k(\boldsymbol{\alpha})} \right)\!,\quad
\end{eqnarray}
where $\boldsymbol{\mu}=[\lambda_1,\ldots,\lambda_K]^{\mathrm{T}}$ is the dual variable vector introduced for each inequality constraint in \eqref{Pa2_rate}. From the strong duality, $\mathbf{P}_{\alpha 2}$ is equivalently reformulated to a dual problem as follows:
\begin{eqnarray}\label{Pa3_prob}
    \mathbf{P}_{\alpha 3}:
    \underset {\boldsymbol{\mu} \succeq 0}{\text{minimize}} \quad \underset {\boldsymbol{\theta} }{\text{maximize}} \quad L(\boldsymbol{\theta},\boldsymbol{\mu}).
\end{eqnarray}
The optimal solution of $\mu_k$ can then be obtained by evaluating the first-order condition $\partial L(\boldsymbol{\theta},\boldsymbol{\mu})/ \partial \theta_k$ and applying the trivial solution to $\mathbf{P}_{\alpha 2}$ as
\begin{eqnarray}\label{mu_opt}
    \mu_k^{o} = \frac{\psi \hat{B}_k(\boldsymbol{\alpha})}{\hat{A}_k(\boldsymbol{\alpha})+\hat{B}_k(\boldsymbol{\alpha})}.
\end{eqnarray}
Besides, $\mu_k \geq 0$ is automatically satisfied in this case.  Using  \eqref{LD_func_a} and \eqref{mu_opt},  $\mathbf{P}_{\alpha 2}$ is reformulated as
\begin{eqnarray}\label{Pa4_prob}
	\mathbf{P}_{\alpha 4}:
  \underset {\boldsymbol{\theta} }{\text{maximize}} \quad L(\boldsymbol{\theta},\boldsymbol{\mu}^o).
\end{eqnarray}
Further, it can be shown the solution to $\mathbf{P}_{\alpha 4}$ satisfies $\mathbf{P}_{\alpha }$ when combined with the outer optimization over $\boldsymbol{\alpha}$ and after several mathematical interpretations \cite{Shen2018,Guo2019Weighted}. Then, the tags' reflection coefficients, $\boldsymbol{\alpha}$, are obtained by solving feasibility problem over $\boldsymbol{\alpha}$ for fixed $\boldsymbol{\theta}$. The corresponding optimization problem is thus given as
\begin{subequations}\label{Pa5_prob}
    \begin{eqnarray}
        \mathbf{P}_{\alpha 5}:
        \text{find} && \boldsymbol{\alpha}, \\
        \text{subject to} && \theta_k^{o} \leq \frac{\hat{A}_k(\boldsymbol{\alpha})}{\hat{B}_k(\boldsymbol{\alpha})},  \label{Pa1_rate}\\ 
        && (1-\alpha_k) p_t \left\vert  \mathbf{d}_k^{\rm{T}} \mathbf{w} \right\vert^2 \geq p_b', \quad \label{Pa1_tag_EH}\\
        &&0 < \alpha_k < 1. \label{Pa1_alpha_constr}
    \end{eqnarray}
\end{subequations}
Algorithm \ref{AlgoAlpha} provides the proposed approach for optimizing the tags' reflection coefficients.  

\begin{algorithm}[hbt!]
\caption{: Reflection coefficient optimization. }\label{AlgoAlpha}
 {\small{\begin{algorithmic}
        \renewcommand{\algorithmicrequire}{\textbf{Initialization:}}
        \renewcommand{\algorithmicensure}{\textbf{Repeat}}
        \Require Initialize $\boldsymbol{\alpha}$ to a feasible value.
        \Ensure
        \State \textbf{Step 1}: Update $\boldsymbol{\mu}$ by \eqref{mu_opt}.
        \State \textbf{Step 2}: Update $\boldsymbol{\theta}$ by solving $\mathbf{P}_{\alpha 4}$ in \eqref{Pa4_prob}.
        \State \textbf{Step 3}: Update $\boldsymbol{\alpha}$ by solving $\mathbf{P}_{\alpha 5}$ in \eqref{Pa5_prob}.
    \end{algorithmic}
\textbf{Until} the value of the objective function converges.\\
\textbf{Output:} The optimal reflection coefficients $\boldsymbol{\alpha}^o$. }}
\end{algorithm}

\begin{algorithm}[hbt!]
\caption{: Overall algorithm. }\label{AlgoOvrl}
 {\small{\begin{algorithmic}
        \renewcommand{\algorithmicrequire}{\textbf{Initialization:}}
        \renewcommand{\algorithmicensure}{\textbf{Repeat}}
        \Require Initialize $\mathbf{u}_k$ for $k \in \mathcal{K}$ and $\boldsymbol{\alpha}$ to a feasible values.
        \Ensure
        \State \textbf{Step 1}: Update $\mathbf{w}$ by solving $\mathbf{P}_{\mathrm{w}}$ in \eqref{Pw_prob}.
        \State \textbf{Step 2}:  Update $\mathbf{u}_k$ for $k\in \mathcal{K}$ by solving $\mathbf{P}_{\mathrm{u}}$ in \eqref{Pu_prob}.
        \State \textbf{Step 3}:  Update $\boldsymbol{\alpha}$ by solving $\mathbf{P}_{\alpha}$ in \eqref{Pa_prob}.
    \end{algorithmic}
\textbf{Until} the value of the objective function converges.\\
\textbf{Output:} The optimal beamforming $\mathbf{w}^{o}$, receiver combining $\mathbf{u}_k^{o}$, and reflection coefficients $\boldsymbol{\alpha}^o$. }}
\end{algorithm}

\begin{rem}\label{Remark:overal}
Algorithm \ref{AlgoOvrl} presents the overall algorithm for solving the proposed optimization problem, $\mathbf{P1}$ \eqref{P1_prob}. We begin by setting $\mathbf{u}_k$ for $k \in \mathcal{K}$ and $\boldsymbol{\alpha}$ to feasible values. Then, in each iteration, algorithms update better solutions for $\mathbf{w}$, $\mathbf{u}_k$, and $\boldsymbol{\alpha}$ until the objective function no longer improves. The procedure is terminated when the normalized objective function increment is less than $\epsilon=10^{-3}$.
\end{rem}

\subsection{Computational Complexity}
The proposed AO solution is a multi-stage iterative algorithm. Here, the outer loop has three sub-problems for optimizing $\mathbf{w}$, $\mathbf{u}_k$ for $k\in \mathcal{K}$, and $\boldsymbol{\alpha}$. Each sub-problem requires an iterative updating method to solve. Specifically, the computational complexity of Algorithm \ref{Algow} lies in Step 2. Matlab CVX uses SDPT3 solver for solving the sub-problem for $\mathbf{w}$. Thus, the computational complexities of Algorithm Algorithm \ref{Algow} is $\mathcal{O}(M^3K^3)$ \cite{Ben2001book,Borrero2017}. The computational complexity of the sub-problem for solving $\mathbf{u}_k$ for $k\in \mathcal{K}$ is in the matrix inversion, multiplication, and eigenvalue decomposition  \eqref{eqn_optml_u}. Hence, the computational complexity is $\mathcal{O}(K^3(2K^2+1))$. The computational complexity of Algorithm \ref{AlgoAlpha}  lies in step 3. Similar to Algorithm \ref{Algow}, CVX Matlab uses an SDPT3 solver to handle this optimization problem. Therefore, the computational complexity of Algorithm \ref{AlgoAlpha} is $\mathcal{O}(M^3K^3)$ \cite{Ben2001book,Borrero2017}. Hence, the total complexity of the proposed AO solution is $\mathcal{O}(I_o (I_w M^3K^3 + K^3(2K^2+1) + I_{\alpha} M^3K^3))$, where $I_w$, $I_{\alpha}$, and $I_o$ are the iteration numbers of Algorithm \ref{Algow}, Algorithm \ref{AlgoAlpha}, and the overall algorithm (outer loop -- Algorithm \ref{AlgoOvrl}), respectively \cite{Ben2001book,Borrero2017}.

\subsection{Algorithm Convergence}

As a well-established concept, the overall algorithm convergence in AO is guaranteed as long as the individual blocks within the AO algorithm converge \cite{Bezdek2003}. 

In our proposed solution, we utilize the FP technique to solve $\mathbf{w}$ and $\boldsymbol{\alpha}$ blocks, whereas $\mathbf{u}_k$ for $k \in \mathcal{K}$ is obtained as a closed-form solution applying the Rayleigh ratio quotient approach. Hence, overall algorithm convergence relies on the FP. Interestingly, the FP yields a fixed-point iteration method with provable convergence \cite{Kaiming2022thesis,  Borrero2017, Shen2018}, ensuring the convergence of our proposed AO method. Nonetheless, our simulation results also corroborate the validity of this claim (Fig.~\ref{fig_Obj_Itr}).

\section{Simulation Results}\label{Sec_sim}
Herein, we present simulation examples for evaluating the performance of the proposed cell-free \bbc system. 

We adopt a three-slope model for modeling the large-scale fading $\zeta_v$ \eqref{eqn_channel_model} with operating frequency, $f_c$, in \qty{}{\MHz} \cite{Ngo2017}. In particular, for $d_0=\qty{10}{\m}$ and $d_1=\qty{50}{\m}$, the path-loss exponent is (i) \num{3.5} if the distance between two nodes in \qty{}{\m} (denoted by $d$) is larger than $d_1$, (ii) \num{2} if $d_1\geq d > d_0$, and (iii) \num{0} if $d \leq d_0$. When $d> d_1$, the Hata-COST231 propagation model is employed. The path-loss in \qty{}{\dB} is then defined as
\begin{eqnarray}
    \zeta_v  \!=\! \begin{cases}
    \!-L-35\log_{10}(d), \,\, \text{if}\,\, d>d_1, \\
    \!-L-15\log_{10}(d_1)-20\log_{10}(d), \,\, \text{if}\,\, d_0<d\leq d_1,\\
    \!-L-15\log_{10}(d_1)-20\log_{10}(d_0), \,\, \text{if}\,\, d\leq d_0,
    \end{cases} 
\end{eqnarray}
where $L=46.3 + 33.9 \log_{10}(f_c)- 13.82\log_{10}(h_t) - (1.1\log_{10}(f_c)-0.7)h_r + (1.56 \log_{10}(f_c) - 0.8)$. Here, $h_t$ and $h_r$ denote the transmitter and receiver antenna heights in \qty{}{\m}, respectively, i.e., AP antenna height, $h_{AP}$, tag antenna height, $h_{T}$, and reader antenna height, $h_{R}$. The AWGN variance, $\sigma^2$ is modeled as $\sigma^2= 10 \log_{10}{(N_0 B N_f)}$\,dBm, where $N_0=-174$\,dBm/Hz, $B$ is the bandwidth, and $N_f$ is the noise figure. 

To model the coverage area, e.g., warehouse, we consider a $100\times 100$\,\qty{}{\m^2} square area, the reader is located at the center, the APs are uniformly distributed, and the tags are randomly distributed within the area. Unless otherwise specified, Table \ref{tab_sim_para} gives the simulation parameters.

\begin{table}
\caption{Simulation settings.} 
\vspace{-0mm}
\label{tab_sim_para}
\centering 
\begin{tabular}{c c c c} 
\hline 
    Parameter & Value & Parameter & Value \\ [0.5ex] 
    \hline \hline
    $f_c$ & \qty{2}{\GHz}  &$M$ & \num{36} \\
    $B$ & \qty{10}{\MHz}  &$L$ & \num{4} \\ 
    $N_f$ & \qty{10}{\dB} & $K$ & \{\num{3}, \num{5}\}\\
    $d_0,d_1$ & \{10, 50\}\qty{}{\m} & $p_b$ & \qty{-20}{\dB m}   \\
    $h_{AP},h_{T}, h_{R}$ & \{15, 1, 1.6\}\qty{}{\m} & $p_p$ &  \qty{20}{\dB m}  \\
    $\tau_c$ & \num{1000} & $\tau_p$ &  \num{5}  \\
\hline 
\end{tabular}
\vspace{-0mm}
\end{table}

\subsection{Channel Estimation}
We first investigate the performance of our channel estimation method (section \ref{channel_estimate}), considering the LS estimator. 
 
The quality of the channel estimator is  assessed in terms of   normalized MSE, which  is defined as
\begin{eqnarray}
    \text{Normalized MSE} = \frac{\mathbb{E} \left\{  \left\Vert  \mathbf{v} - \hat{\mathbf{v}} \right \Vert^2 \right\}} {\mathbb{E} \left\{  \left\Vert  \mathbf{v} \right \Vert^2 \right\}}, 
\end{eqnarray}
where $\mathbf{v} \in \{\mathbf{h}_{0,m}, \mathbf{h}_{k,m}, f_{k,m} \}$.

\begin{figure}[!t]\vspace{-0mm}	
\centering
\fontsize{14}{14}\selectfont 
    \resizebox{.585\totalheight}{!}{\input{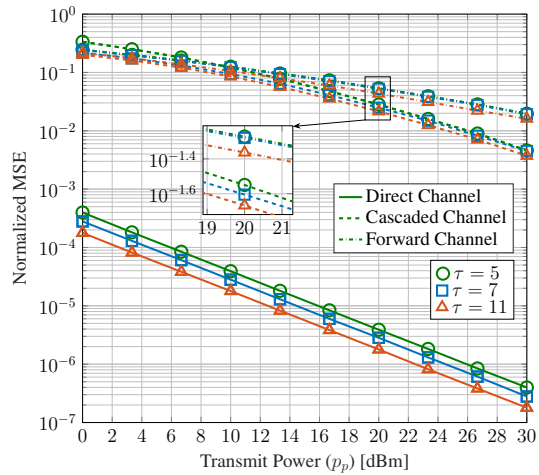}}\vspace{-0mm}
    \caption{Normalized MSE of direct, cascaded, and forward channels for $K=3$, and $\tau=\{5,7,11\}$.}
	\label{fig_NMSE_Ptx} \vspace{-0mm}
\end{figure}

Fig. \ref{fig_NMSE_Ptx} shows the normalized MSE performance of the LS estimator versus the per-AP pilot transmit power $p_p$ for the direct, cascaded, and forward channels, considering $K=3$, and $\tau = \{5,7,11\}$. Our method demonstrates high accuracy in estimating all the channels for multiple tag scenarios. However, the cascaded and the forward channel normalized MSE have high values when compared to the direct channel normalized MSE. This is primarily due to the double path loss present in the cascaded and the forward channels and the tag’s reflection coefficient $\alpha$. We also observe that increasing the pilot sequence length improves the estimation accuracy and lowers the required transmit power for a given  MSE level. For instance, for the direct link, a normalized MSE of $10^{-5}$ is achieved at \qty{12}{\dB m} and \qty{16}{\dB m} for  $\tau =11$ and $\tau =5$, respectively. Similarly, increasing the pilot length from $\tau =5$ and $\tau =11$ can save \qty{2.5}{\dB m} dB and \qty{2}{\dB m} of transmit power, respectively, to obtain a normalized MSE of $2\times 10^{-2}$ in the forward channel and $5\times 10^{-3}$ in the cascaded channel.

Note that using the highly accurate estimates of the direct link channels, the reader is able to effectively cancel out the DLI, thereby justifying the assumption stated in Remark \ref{rem_direct_channel}.
 
In the following, we use these channel estimates to evaluate the performance of our proposed schemes for the \bbc network, for $p_p = \qty{20}{\dB m}$, and $\tau = 5$.

\subsection{Convergence Rate of the Proposed Algorithm}
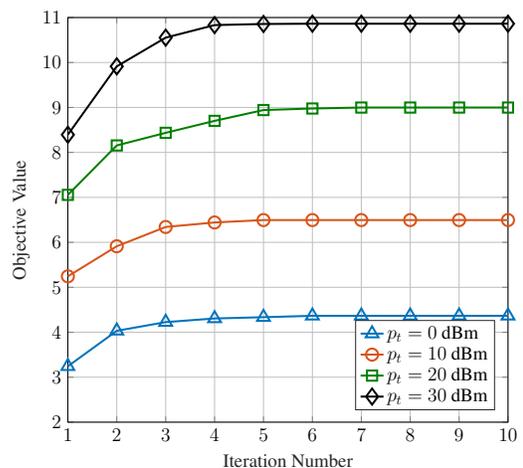
\begin{figure}[!t]\vspace{-0mm}	
\centering
\fontsize{14}{14}\selectfont 
    \resizebox{.58\totalheight}{!}{
%
%
\definecolor{mycolor1}{rgb}{0.00000,0.45000,0.74000}%
\definecolor{mycolor2}{rgb}{0.85000,0.33000,0.10000}%
\begin{tikzpicture}

\begin{axis}[%
width=4.54in,
height=4.177in,
at={(0.762in,0.564in)},
scale only axis,
xmin=1,
xmax=10,
xlabel style={font=\color{white!15!black}},
xlabel={Iteration Number},
ymin=2,
ymax=11,
ylabel style={font=\color{white!15!black}},
ylabel={Objective Value},
axis background/.style={fill=white},
xmajorgrids,
ymajorgrids,
legend style={at={(0.97,0.03)}, anchor=south east, legend cell align=left, align=left, draw=white!15!black}
]
\addplot [color=mycolor1, line width=1.5pt, mark size=5.5pt, mark=triangle, mark options={solid, mycolor1}]
  table[row sep=crcr]{%
1	3.24288587928425\\
2	4.0318868187222\\
3	4.22521656514143\\
4	4.3063190216339\\
5	4.3348283550424\\
6	4.3657614190971\\
7	4.3657614190971\\
8	4.3657614190971\\
9	4.3657614190971\\
10	4.3657614190971\\
11	4.3657614190971\\
12	4.3657614190971\\
13	4.3657614190971\\
14	4.3657614190971\\
15	4.3657614190971\\
16	4.3657614190971\\
17	4.3657614190971\\
18	4.3657614190971\\
19	4.3657614190971\\
20	4.3657614190971\\
};
\addlegendentry{$p_t=0$ dBm}

\addplot [color=mycolor2, line width=1.5pt, mark size=4.5pt, mark=o, mark options={solid, mycolor2}]
  table[row sep=crcr]{%
1	5.24340822271533\\
2	5.91260679963182\\
3	6.34170657899472\\
4	6.44201360234546\\
5	6.49534595415446\\
6	6.49534595415446\\
7	6.49534595415446\\
8	6.49534595415446\\
9	6.49534595415446\\
10	6.49534595415446\\
11	6.49534595415446\\
12	6.49534595415446\\
13	6.49534595415446\\
14	6.49534595415446\\
15	6.49534595415446\\
16	6.49534595415446\\
17	6.49534595415446\\
18	6.49534595415446\\
19	6.49534595415446\\
20	6.49534595415446\\
};
\addlegendentry{$p_t=10$ dBm}

\addplot [color=black!50!green, line width=1.5pt, mark size=3.8pt, mark=square, mark options={solid, black!50!green}]
  table[row sep=crcr]{%
1	7.05548076946311\\
2	8.15477231130358\\
3	8.43572088817519\\
4	8.70143455886062\\
5	8.94226720129299\\
6	8.97651962920024\\
7	8.99786239393574\\
8	8.99786239393574\\
9	8.99786239393574\\
10	8.99786239393574\\
11	8.99786239393574\\
12	8.99786239393574\\
13	8.99786239393574\\
14	8.99786239393574\\
15	8.99786239393574\\
16	8.99786239393574\\
17	8.99786239393574\\
18	8.99786239393574\\
19	8.99786239393574\\
20	8.99786239393574\\
};
\addlegendentry{$p_t=20$ dBm}

\addplot [color=black, line width=1.5pt, mark size=6.0pt, mark=diamond, mark options={solid, black}]
  table[row sep=crcr]{%
1	8.39400693683863\\
2	9.91385665972427\\
3	10.552663638391\\
4	10.8320910030995\\
5	10.8567213303363\\
6	10.8627378708171\\
7	10.8627378708171\\
8	10.8627378708171\\
9	10.8627378708171\\
10	10.8627378708171\\
11	10.8627378708171\\
12	10.8627378708171\\
13	10.8627378708171\\
14	10.8627378708171\\
15	10.8627378708171\\
16	10.8627378708171\\
17	10.8627378708171\\
18	10.8627378708171\\
19	10.8627378708171\\
20	10.8627378708171\\
};
\addlegendentry{$p_t=30$ dBm}

\end{axis}

\begin{axis}[%
width=5.858in,
height=5.125in,
at={(0in,0in)},
scale only axis,
xmin=0,
xmax=1,
ymin=0,
ymax=1,
axis line style={draw=none},
ticks=none,
axis x line*=bottom,
axis y line*=left
]
\end{axis}
\end{tikzpicture}%
}\vspace{-0mm}
    \caption{The convergence of the objective value of the overall algorithm.}
	\label{fig_Obj_Itr} \vspace{-0mm}
\end{figure}

Before proceeding to the performance analysis, we investigate the convergence rate of the proposed algorithm (Section \ref{Sec:Proposed_scheme}, Remark \ref{Remark:overal}). 
To this end, Fig. \ref{fig_Obj_Itr}  shows its convergence behavior for when the AP is at  $p_t=\{0,10,20,30\}\,\text{dBm}$. The objective function of the overall algorithm is the sum rate. The stopping condition for convergence is that the increment of the normalized objective function is less than $\epsilon=10^{-3}$. As shown in Fig. \ref{fig_Obj_Itr}, the sum rate achieved by the overall  Algorithm \ref{AlgoOvrl} increases rapidly and saturates as the number of iterations increases. Specifically, it  converges in less than five  iterations regardless of the AP transmit power. 

\subsection{Performance Analysis}
In this section, we assess the performance of the proposed cell-free \bbc network. We evaluate our power allocation  and channel estimation schemes  in terms their ability to enhance the received power at the tags and  the sum rate. We consider two benchmarks alongside our proposed scheme: 
\begin{itemize}
    \item[$A_1$] \textit{Random beamforming}: Optimal beamforming in a multi-antenna communication system relies on CSI. When CSI is unavailable, random beamforming may be used as an alternative, which might not seem ideal at first. To assess the effectiveness of random beamforming, we may compare it with optimal beamforming, assuming perfect CSI, in a system with $M$ transmit antennas and $n$ users. Surprisingly, in this scenario, both optimal and random beamforming achieve the same asymptotic sum-rate capacity of $M \log (n)$ \cite{Jaehak2003, Sharif2003}. This observation justifies the use of random beamforming in cases where CSI is not available. To implement random beamforming, we choose complex Gaussian random vectors for $\mathbf{w}$ and $\mathbf{u}_k, (k \in \mathcal{K})$, satisfying the per-AP transmit power constraint \eqref{P1_AP_power} and the reader's normalized power constraint \eqref{P1_detector}, respectively. The advantage of this approach is that it requires no CSI. Additionally, the reflection coefficients are set to \num{0.6} in this specific case.
    
    \item[$A_2$]\textit{Optimal beamforming with perfect CSI}: $\mathbf{w}$,  $\mathbf{u}_k\,(k\in \mathcal{K})$, and $\boldsymbol{\alpha}$ are obtained through the proposed optimization framework with perfect CSI, i.e., without performing channel estimation. 

    \item[$A_3$]\textit{Optimal beamforming with estimated CSI}: $\mathbf{w}$,  $\mathbf{u}_k\,(k\in \mathcal{K})$, and $\boldsymbol{\alpha}$ are obtained using the proposed optimization framework with estimated CSI using the proposed channel estimation. 
\end{itemize}

\subsubsection{Received Power at the Tag} 
Increasing this quantity is critical for ensuring that the tag can be activated.  

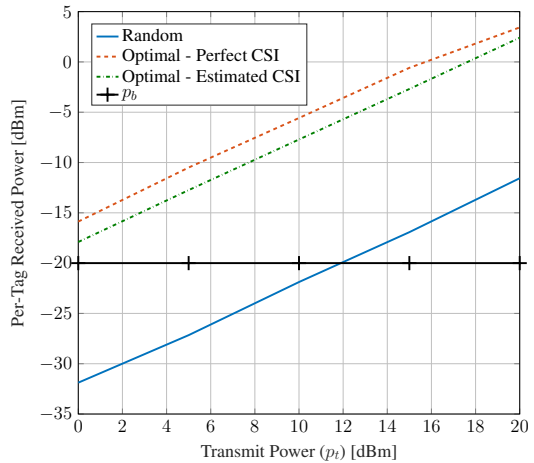
\begin{figure}[!t]\vspace{-0mm}	
\centering
\fontsize{14}{14}\selectfont 
    \resizebox{.56\totalheight}{!}{
%
%
\definecolor{mycolor1}{rgb}{0.00000,0.44700,0.74100}%
\definecolor{mycolor2}{rgb}{0.85000,0.32500,0.09800}%
\definecolor{mycolor3}{rgb}{0.00000,0.49804,0.00000}%
\begin{tikzpicture}

\begin{axis}[%
width=4.755in,
height=4.338in,
at={(0.798in,0.586in)},
scale only axis,
xmin=0,
xmax=20,
xlabel style={font=\color{white!15!black}},
xlabel={Transmit Power ($p_t$) [dBm]},
ymin=-35,
ymax=5,
ylabel style={font=\color{white!15!black}},
ylabel={Per-Tag Received Power [dBm]},
axis background/.style={fill=white},
xmajorgrids,
ymajorgrids,
legend style={at={(0.03,0.97)}, anchor=north west, legend cell align=left, align=left, draw=white!15!black}
]
\addplot [color=mycolor1, line width=1.5pt]
  table[row sep=crcr]{%
0	-31.8826195430274\\
5	-27.1604786432653\\
10	-21.8801855046459\\
15	-16.923011295875\\
20	-11.5621141080266\\
};
\addlegendentry{Random}

\addplot [color=mycolor2, dashed, line width=1.5pt]
  table[row sep=crcr]{%
0	-15.8705031447201\\
5	-10.4996922686381\\
10	-5.57277136833047\\
15	-0.589061302701424\\
20	3.4294364760625\\
};
\addlegendentry{Optimal - Perfect CSI}

\addplot [color=mycolor3, dashdotted, line width=1.5pt]
  table[row sep=crcr]{%
0	-17.8980061240683\\
5	-12.7426060913692\\
10	-7.70964402799715\\
15	-2.6912164646192\\
20	2.40710048611543\\
};
\addlegendentry{Optimal - Estimated CSI}

\addplot [color=black, line width=1.5pt, mark size=5.5pt, mark=+, mark options={solid, black}]
  table[row sep=crcr]{%
0	-20\\
5	-20\\
10	-20\\
15	-20\\
20	-20\\
};
\addlegendentry{$p_b$}

\end{axis}

\begin{axis}[%
width=6.135in,
height=5.323in,
at={(0in,0in)},
scale only axis,
xmin=0,
xmax=1,
ymin=0,
ymax=1,
axis line style={draw=none},
ticks=none,
axis x line*=bottom,
axis y line*=left
]
\end{axis}

\end{tikzpicture}%
}\vspace{-0mm}
    \caption{Per-tag received power versus the transmit power.}
	\label{fig_RxPower_Ptx} \vspace{-0mm}
\end{figure}

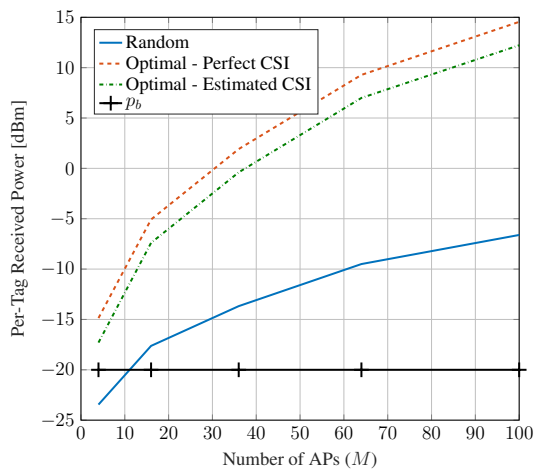
\begin{figure}[!t]\vspace{-0mm}	
\centering
\fontsize{14}{14}\selectfont 
    \resizebox{.58\totalheight}{!}{
%
%
\definecolor{mycolor1}{rgb}{0.00000,0.44700,0.74100}%
\definecolor{mycolor2}{rgb}{0.85000,0.32500,0.09800}%
\definecolor{mycolor3}{rgb}{0.00000,0.49804,0.00000}%
\begin{tikzpicture}

\begin{axis}[%
width=4.54in,
height=4.177in,
at={(0.762in,0.564in)},
scale only axis,
xmin=0,
xmax=100,
xlabel style={font=\color{white!15!black}},
xlabel={Number of APs ($M$)},
ymin=-25,
ymax=15,
ylabel style={font=\color{white!15!black}},
ylabel={Per-Tag Received Power [dBm]},
axis background/.style={fill=white},
xmajorgrids,
ymajorgrids,
legend style={at={(0.03,0.97)}, anchor=north west, legend cell align=left, align=left, draw=white!15!black}
]
\addplot [color=mycolor1, line width=1.5pt]
  table[row sep=crcr]{%
4	-23.4533040003347\\
16	-17.6257845828747\\
36	-13.6780959091625\\
64	-9.50001017753453\\
100	-6.61172777985891\\
};
\addlegendentry{Random}

\addplot [color=mycolor2, dashed, line width=1.5pt]
  table[row sep=crcr]{%
4	-14.8633695516919\\
16	-5.07386056566315\\
36	1.91876851026278\\
64	9.28061505005466\\
100	14.5480385975106\\
};
\addlegendentry{Optimal - Perfect CSI}

\addplot [color=mycolor3, dashdotted, line width=1.5pt]
  table[row sep=crcr]{%
4	-17.2978426123861\\
16	-7.39488424081358\\
36	-0.391771871091704\\
64	6.99270162187204\\
100	12.2216895629161\\
};
\addlegendentry{Optimal - Estimated CSI}

\addplot [color=black, line width=1.5pt, mark size=5.5pt, mark=+, mark options={solid, black}]
  table[row sep=crcr]{%
4	-20\\
16	-20\\
36	-20\\
64	-20\\
100	-20\\
};
\addlegendentry{$p_b$}

\end{axis}

\begin{axis}[%
width=5.858in,
height=5.125in,
at={(0in,0in)},
scale only axis,
xmin=0,
xmax=1,
ymin=0,
ymax=1,
axis line style={draw=none},
ticks=none,
axis x line*=bottom,
axis y line*=left
]
\end{axis}
\end{tikzpicture}%
}\vspace{-0mm}
    \caption{Per-tag received power versus the number of APs for $p_t=\qty{20}{\dB m}$.}
	\label{fig_RxPower_M} \vspace{-0mm}
\end{figure}

To achieve this objective, we present in Fig.~\ref{fig_RxPower_Ptx} and Fig.~\ref{fig_RxPower_M} the graphical representations of the received power experienced by a tag, delineated with respect to variations in transmit power and the number of APs. Notably, these figures also incorporate the activation threshold, $p_b=\qty{-20}{\dB m}$, a reference point derived from established parameters for commercial RFID tags \cite{GAOtags}, thereby providing invaluable insights into our analysis. Examining Fig.~\ref{fig_RxPower_Ptx} and Fig.~\ref{fig_RxPower_M} underscores the efficacy of our proposed channel estimation and optimization frameworks. Across a spectrum of transmit power settings and AP quantities, our frameworks consistently yield higher power delivery to the tags. In stark contrast, the utilization of random beamforming ($A_1$) necessitates escalated transmit power levels and an increased number of APs to attain the power threshold requisite for the tags. Illustratively, for instance, the $A_1$ approach mandates a minimum of \qty{12}{\dB m}  when paired with \num{36} APs (Fig.~\ref{fig_RxPower_Ptx}), or \num{12} APs at \qty{20}{\dB m}  (Fig.~\ref{fig_RxPower_M}) to exceed the activation threshold.  Evidently, from an energy efficiency perspective, random beamforming exhibits suboptimal performance. Our proposed algorithms clearly bring much-needed energy savings.

However, it is important to note that our proposed channel estimation scheme, while effective, falls short of the ideal performance achieved by perfect CSI benchmark $A_2$. Despite this disparity, there are strategic measures to bridge the gap:
\begin{enumerate}
    \item 
\textbf{Allocating higher transmit power at APs: }By boosting the power at APs, we can enhance the signal-to-noise ratio and mitigate the impact of imperfect channel estimation.

\item \textbf{Deploying more APs:} Increasing the density of APs can provide additional diversity and redundancy, which in turn can help compensate for the performance gap.

\item \textbf{Employing longer pilot sequences:} Longer pilot sequences offer finer granularity in channel estimation, allowing for improved accuracy even with non-ideal schemes.

\end{enumerate}
In essence, while our proposed CSI estimation method may not attain perfection, it can harmoniously integrate with these complementary solutions to attain remarkably high-performance standards.


\subsubsection{Achievable Sum Rate}
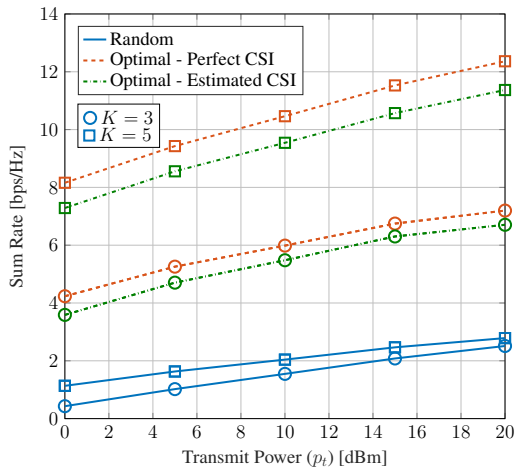
\begin{figure}[!t]\vspace{-0mm}	
\centering
\fontsize{14}{14}\selectfont 
    \resizebox{.58\totalheight}{!}{
%
%
\definecolor{mycolor1}{rgb}{0.00000,0.45000,0.74000}%
\definecolor{mycolor2}{rgb}{0.85000,0.33000,0.10000}%
\begin{tikzpicture}

\begin{axis}[%
width=4.54in,
height=4.177in,
at={(0.762in,0.564in)},
scale only axis,
xmin=0,
xmax=20,
xlabel style={font=\color{white!15!black}},
xlabel={Transmit Power ($p_t$) [dBm]},
ymin=0,
ymax=14,
ylabel style={font=\color{white!15!black}},
ylabel={Sum Rate [bps/Hz]},
axis background/.style={fill=white},
xmajorgrids,
ymajorgrids,
legend style={at={(0.03,0.97)}, anchor=north west, legend cell align=left, align=left, draw=white!15!black}
]
\addplot [color=mycolor1, line width=1.5pt]
  table[row sep=crcr]{%
0	0.429671428655998\\
5	1.02016757718792\\
10	1.54737320857971\\
15	2.08059047943262\\
20	2.51237542321803\\
};
\addlegendentry{Random}

\addplot [color=mycolor1, line width=1.5pt, only marks, mark size=4.5pt, mark=o, mark options={solid, mycolor1}, forget plot]
  table[row sep=crcr]{%
0	0.429671428655998\\
5	1.02016757718792\\
10	1.54737320857971\\
15	2.08059047943262\\
20	2.51237542321803\\
};\label{f5_p1}
\addplot [color=mycolor2, dashed, line width=1.5pt]
  table[row sep=crcr]{%
0	4.23392536381007\\
5	5.25701208984282\\
10	5.98544011313827\\
15	6.74774827572697\\
20	7.19698882882984\\
};
\addlegendentry{Optimal - Perfect CSI}

\addplot [color=mycolor2, dashed, line width=1.5pt, mark size=4.5pt, mark=o, mark options={solid, mycolor2}, forget plot]
  table[row sep=crcr]{%
0	4.23392536381007\\
5	5.25701208984282\\
10	5.98544011313827\\
15	6.74774827572697\\
20	7.19698882882984\\
};
\addplot [color=black!50!green, dashdotted, line width=1.5pt]
  table[row sep=crcr]{%
0	3.58823939647528\\
5	4.70401218275935\\
10	5.47898468312612\\
15	6.30171673839563\\
20	6.70610327576233\\
};
\addlegendentry{Optimal - Estimated CSI}

\addplot [color=black!50!green, dashdotted, line width=1.5pt, mark size=4.5pt, mark=o, mark options={solid, black!50!green}, forget plot]
  table[row sep=crcr]{%
0	3.58823939647528\\
5	4.70401218275935\\
10	5.47898468312612\\
15	6.30171673839563\\
20	6.70610327576233\\
};
\addplot [color=mycolor1, line width=1.5pt, mark size=3.8pt, mark=square, mark options={solid, mycolor1}, forget plot]
  table[row sep=crcr]{%
0	1.13456704337671\\
5	1.63084044487262\\
10	2.04114335878855\\
15	2.46571339767999\\
20	2.78564385618721\\
};
\addplot [color=mycolor1, line width=1.5pt, only marks, mark size=3.8pt, mark=square, mark options={solid, mycolor1}, forget plot]
  table[row sep=crcr]{%
0	1.13456704337671\\
5	1.63084044487262\\
10	2.04114335878855\\
15	2.46571339767999\\
20	2.78564385618721\\
};\label{f5_p2}
\addplot [color=mycolor2, dashed, line width=1.5pt, mark size=3.8pt, mark=square, mark options={solid, mycolor2}, forget plot]
  table[row sep=crcr]{%
0	8.1595416532115\\
5	9.42850011519279\\
10	10.4630430285089\\
15	11.5276593646268\\
20	12.3636936570747\\
};
\addplot [color=black!50!green, dashdotted, line width=1.5pt, mark size=3.8pt, mark=square, mark options={solid, black!50!green}, forget plot]
  table[row sep=crcr]{%
0	7.2862517355497\\
5	8.55619240455971\\
10	9.54262416427327\\
15	10.5697617161833\\
20	11.3687484789175\\
};
\end{axis}

\begin{axis}[%
width=5.858in,
height=5.125in,
at={(0in,0in)},
scale only axis,
xmin=0,
xmax=1,
ymin=0,
ymax=1,
axis line style={draw=none},
ticks=none,
axis x line*=bottom,
axis y line*=left
]
\end{axis}

\node [draw,fill=white] at (rel axis cs: 0.225,0.7) {\shortstack[l]{
\ref{f5_p1} $K=3$ \\
\ref{f5_p2} $K=5$}};

\end{tikzpicture}%
}\vspace{-0mm}
    \caption{Sum rate versus the transmit power.}
	\label{fig_SumRate_Ptx} \vspace{-0mm}
\end{figure}

\begin{figure}[!t]\vspace{-0mm}	
\centering
\fontsize{14}{14}\selectfont 
    \resizebox{.58\totalheight}{!}{
%
%
\definecolor{mycolor1}{rgb}{0.00000,0.45000,0.74000}%
\definecolor{mycolor2}{rgb}{0.85000,0.33000,0.10000}%
\begin{tikzpicture}

\begin{axis}[%
width=4.54in,
height=4.177in,
at={(0.762in,0.564in)},
scale only axis,
xmin=0,
xmax=100,
xlabel style={font=\color{white!15!black}},
xlabel={Number of APs ($M$)},
ymin=0,
ymax=20,
ylabel style={font=\color{white!15!black}},
ylabel={Sum Rate [bps/Hz]},
axis background/.style={fill=white},
xmajorgrids,
ymajorgrids,
legend style={at={(0.03,0.97)}, anchor=north west, legend cell align=left, align=left, draw=white!15!black}
]
\addplot [color=mycolor1, line width=1.5pt]
  table[row sep=crcr]{%
4	1.20246561431016\\
16	1.59849943147678\\
36	1.8803004659678\\
64	2.18361339466253\\
100	2.32962926349482\\
};
\addlegendentry{Random}

\addplot [color=mycolor1, line width=1.5pt, only marks, mark size=4.5pt, mark=o, mark options={solid, mycolor1}, forget plot]
  table[row sep=crcr]{%
4	1.20246561431016\\
16	1.59849943147678\\
36	1.8803004659678\\
64	2.18361339466253\\
100	2.32962926349482\\
};\label{f7_p1}
\addplot [color=mycolor2, dashed, line width=1.5pt]
  table[row sep=crcr]{%
4	4.7224541547522\\
16	6.58394198216456\\
36	8.42023159208595\\
64	10.1929110500411\\
100	12.3118583454025\\
};
\addlegendentry{Optimal - Perfect CSI}

\addplot [color=mycolor2, dashed, line width=1.5pt, mark size=4.5pt, mark=o, mark options={solid, mycolor2}, forget plot]
  table[row sep=crcr]{%
4	4.7224541547522\\
16	6.58394198216456\\
36	8.42023159208595\\
64	10.1929110500411\\
100	12.3118583454025\\
};
\addplot [color=black!50!green, dashdotted, line width=1.5pt]
  table[row sep=crcr]{%
4	3.37741517132905\\
16	5.2540651679083\\
36	7.22544226785734\\
64	9.14121421029744\\
100	11.3743584874012\\
};
\addlegendentry{Optimal - Estimated CSI}

\addplot [color=black!50!green, dashdotted, line width=1.5pt, mark size=4.5pt, mark=o, mark options={solid, black!50!green}, forget plot]
  table[row sep=crcr]{%
4	3.37741517132905\\
16	5.2540651679083\\
36	7.22544226785734\\
64	9.14121421029744\\
100	11.3743584874012\\
};
\addplot [color=mycolor1, line width=1.5pt, mark size=3.8pt, mark=square, mark options={solid, mycolor1}, forget plot]
  table[row sep=crcr]{%
4	1.67744343287352\\
16	1.98972376528346\\
36	2.23305979188325\\
64	2.49886873752408\\
100	2.60784271899346\\
};
\addplot [color=mycolor1, line width=1.5pt, only marks, mark size=3.8pt, mark=square, mark options={solid, mycolor1}, forget plot]
  table[row sep=crcr]{%
4	1.67744343287352\\
16	1.98972376528346\\
36	2.23305979188325\\
64	2.49886873752408\\
100	2.60784271899346\\
};\label{f7_p2}
\addplot [color=mycolor2, dashed, line width=1.5pt, mark size=3.8pt, mark=square, mark options={solid, mycolor2}, forget plot]
  table[row sep=crcr]{%
4	9.4703803681175\\
16	11.9647334946645\\
36	14.1448811106227\\
64	16.3891951128107\\
100	18.303918662271\\
};
\addplot [color=black!50!green, dashdotted, line width=1.5pt, mark size=3.8pt, mark=square, mark options={solid, black!50!green}, forget plot]
  table[row sep=crcr]{%
4	8.23402612768561\\
16	10.0151942051864\\
36	12.2563674229966\\
64	14.4802277280984\\
100	17.1000702207341\\
};
\end{axis}

\begin{axis}[%
width=5.858in,
height=5.125in,
at={(0in,0in)},
scale only axis,
xmin=0,
xmax=1,
ymin=0,
ymax=1,
axis line style={draw=none},
ticks=none,
axis x line*=bottom,
axis y line*=left
]
\end{axis}

\node [draw,fill=white] at (rel axis cs: 0.225,0.7) {\shortstack[l]{
\ref{f7_p1} $K=3$ \\
\ref{f7_p2} $K=5$}};
\end{tikzpicture}%
}\vspace{-0mm}
    \caption{Sum rate versus the number of APs for $p_t=\qty{20}{\dB m}$.}
	\label{fig_SumRate_M} \vspace{-0mm}
\end{figure}
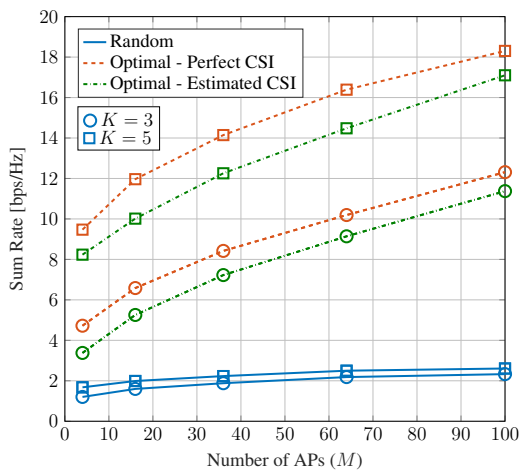

The achievable sum rate of the tags is investigated in Fig.~\ref{fig_SumRate_Ptx} and Fig.~\ref{fig_SumRate_M}, for different numbers of tags $K = \{3,5\}$.

Fig. \ref{fig_SumRate_Ptx} illustrates the sum rate achieved by the tags as a function of the transmit power, $p_t$. It reveals that random beamforming designs, which do not rely on CSI, attain the lowest sum rate as they shape the beam in arbitrary directions.  In contrast, our optimal beamforming designs achieve significant sum rates under both perfect and estimated CSI. Additionally, the achieved sum rate using the estimated CSI is comparable to that achieved with perfect CSI, demonstrating the efficiency of our channel estimation strategy. In particular, the optimal designs under perfect and estimated CSI improve the achieved sum rate respectively \qty{425}{\percent}  and \qty{375}{\percent}  compared to the random designs for $K=5$, and $p_t = \qty{10}{\dB m}.$

Fig. \ref{fig_SumRate_M} also depicts the achieved sum rate versus the number of APs for $p_t = \qty{20}{\dB m}$, and $K = \{3,5 \}$.   As observed, unlike random designs, through optimal beamforming designs, the achieved sum rate significantly improves due to the efficient use of resources. These findings reveal the effectiveness of our proposed resource allocation and CSI estimation frameworks.

\subsubsection{Tags with Fixed and Reconfigurable $\boldsymbol{\alpha}$}

\begin{figure}[!t]\vspace{-0mm}	
\centering
\fontsize{14}{14}\selectfont 
    \resizebox{.58\totalheight}{!}{
%
%
\definecolor{mycolor1}{rgb}{0.00000,0.44700,0.74100}%
\definecolor{mycolor2}{rgb}{0.85000,0.32500,0.09800}%
\begin{tikzpicture}

\begin{axis}[%
width=4.54in,
height=4.177in,
at={(0.762in,0.564in)},
scale only axis,
xmin=0,
xmax=20,
xlabel style={font=\color{white!15!black}},
xlabel={Transmit Power ($p_t$) [dBm]},
ymin=2,
ymax=10,
ylabel style={font=\color{white!15!black}},
ylabel={Sum Rate [bps/Hz]},
axis background/.style={fill=white},
xmajorgrids,
ymajorgrids,
legend style={legend cell align=left, align=left, draw=white!15!black}
]
\addplot [color=mycolor1, dashed, line width=1.5pt]
  table[row sep=crcr]{%
0	3.80225810017327\\
5	5.04722081428125\\
10	6.03570286817996\\
15	7.0534832191713\\
20	7.87445868516085\\
};
\addlegendentry{$\boldsymbol{\alpha}^o$}

\addplot [color=mycolor2, line width=1.5pt]
  table[row sep=crcr]{%
0	3.12767359778153\\
5	4.29588866514489\\
10	5.2476222113325\\
15	6.23742056169031\\
20	6.90064187386877\\
};
\addlegendentry{$\boldsymbol{\alpha}=0.6$}

\end{axis}

\begin{axis}[%
width=5.858in,
height=5.125in,
at={(0in,0in)},
scale only axis,
xmin=0,
xmax=1,
ymin=0,
ymax=1,
axis line style={draw=none},
ticks=none,
axis x line*=bottom,
axis y line*=left
]
\end{axis}
\end{tikzpicture}%
}\vspace{-0mm}
    \caption{Sum rate versus the transmit power for $K=3$.}
	\label{fig_Rate_alpha_ptx} \vspace{-0mm}
\end{figure}
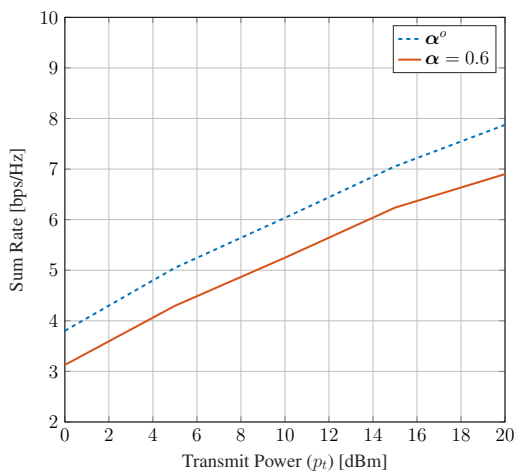
It is interesting to see how much performance gains are possible by going fixed to variable tags. To this end,  we compare the performances of tags with fixed and variable reflection coefficients. Thus, we evaluate fixed tags ($\alpha_k=0.6$ for $k\in \mathcal{K}$)  using the proposed beamforming at APs and combining at the reader. We thus plot the achieved rates of the sum rate as a function of the AP transmit power in Fig.~\ref{fig_Rate_alpha_ptx}. As expected, when the tags use optimal $\boldsymbol{\alpha}^o$, the sum rate significantly improves. For example, the total sum rate achieves by the optimal $\boldsymbol{\alpha}^o$ is $\sim$\qty{1}{bps/\Hz} at $p_t = \qty{20}{\dB m}$ 
than that of the tags with fixed $\boldsymbol{\alpha}$. Furthermore, depending on the application requirements and environment, our optimization framework is flexible enough to handle either  tags with fixed $\alpha$ or tags with variable $\alpha.$ 

\section{Conclusion}\label{Sec_concl}
We investigated the challenge of ensuring robust energy harvests for tags within a broad-ranging \bbc network, spanning extensive areas such as warehouses. Our proposed solution introduces a cell-free, distributed Access AP-assisted \bbc configuration, meticulously engineered to enhance EH reliability across multiple tags, thereby reducing their performance reliance on spatial positioning.

To achieve this objective, we introduced a channel estimation scheme tailored to accommodate the passive transmission characteristics of tags. Leveraging these channel estimates, we determined  optimal beamforming weights at the APs, reflection coefficients at the tags, and combining vectors at the reader. Our aim was to maximize the aggregate tag data rate while satisfying the essential minimum energy requirements.

Notably, the optimization problems posed by these intricate scenarios are inherently non-convex. To address this challenge, we developed solution methods grounded in the Alternating Optimization, Fractional Programming, and Rayleigh Ratio Quotient approaches.

We emphasize that while our study was primarily developed with reconfigurable backscatter tags in mind, it is readily applicable to tags with fixed reflection coefficients. The latter option, characterized by its cost-effectiveness and simplicity, presents a viable networking alternative. Conversely, reconfigurable tags, while more costly, possess the potential to significantly enhance overall network performance.

In summary, our research lays the groundwork for the evolution of cell-free \bbc systems equipped to accommodate multiple tags, thus bolstering their readiness to support emerging  IoT networks.

\bibliographystyle{ieeetr}
\bibliography{ref}

\end{document}